\documentclass[aps,prc,nofootinbib,preprintnumbers,12pt,tightenlines,superscriptaddress]{revtex4}

\usepackage{graphicx}
\usepackage{rotating}
\usepackage{latexsym}

\newcommand{\beq}{\begin{equation}}
\newcommand{\eeq}{\end{equation}}
\newcommand{\bqa}{\begin{eqnarray}}
\newcommand{\eqa}{\end{eqnarray}}

\usepackage{hyperref}
\begin{document}

\title{Conformal Relativistic Viscous Hydrodynamics:
Applications to RHIC results at $\sqrt{s_{NN}}=200$ GeV}

\preprint{INT-08-07}
\preprint{NT-UW-08-10}

\author{Matthew Luzum}
\affiliation{
Theoretical Nuclear Physics Group,
Department of Physics, University of Washington,
Box 351560, Seattle, WA 98195-1560}
\author{Paul Romatschke}
\affiliation{
Institute for Nuclear Theory, University of Washington, 
Box 351550, Seattle, WA, 98195}

\begin{abstract}
A new set of equations for 
relativistic viscous hydrodynamics that captures both
weak-coupling and strong-coupling physics to second
order in gradients has been developed recently. 
We apply this framework to bulk physics at RHIC, both
for standard (Glauber-type) as well as for Color-Glass-Condensate
initial conditions and show that the results do not depend
strongly on the values for the second-order transport coefficients.
Results for multiplicity, radial flow
and elliptic flow are presented and we quote the ratio of viscosity over 
entropy density for which our hydrodynamic model is consistent 
with experimental data. For Color-Glass-Condensate
initial conditions, early thermalization does not
seem to be required in order for hydrodynamics to describe 
charged hadron elliptic flow.
\end{abstract}

\maketitle

\section{Introduction}

The experimental program at the
Relativistic Heavy-Ion Collider (RHIC) at Brookhaven has
generated a wealth of data \cite{Adcox:2004mh,Back:2004je,Arsene:2004fa,Adams:2005dq} on QCD matter
at the highest energy densities obtained in the laboratory.
Remarkably, ideal hydrodynamics seems to offer a sensible
description of the experimental data for bulk properties 
(multiplicity, radial and elliptic flow) of low $p_T$ particles for heavy-ion
collisions at RHIC 
\cite{Teaney:2000cw,Huovinen:2001cy,Kolb:2001qz,Hirano:2002ds,Kolb:2002ve}.

Upon closer inspection, however, not all of this success can be attributed
to modeling the system as an ideal fluid.
For instance, the energy density
distribution which is used as an initial condition for the
hydrodynamic equations is customarily chosen such that the output from
the hydrodynamic model matches the experimental data for the 
multiplicity. Furthermore, the time where the hydrodynamic
model is initialized as well as the temperature (or energy
density) at which the hydrodynamic evolution is stopped
are typically chosen such that the model output matches the 
experimental data for the radial flow.
After these parameters have been fixed, only
the good description of experimental data
for the elliptic flow coefficient can be considered
a success for ideal hydrodynamics (in the sense that
it is parameter-free).

In order to make progress and learn more about the properties
of matter created at RHIC, the task is now to both test
and improve this ideal hydrodynamic model.
The obvious framework for this task is dissipative 
hydrodynamics, since it contains ideal
hydrodynamics as the special case when all dissipative 
transport coefficients (such as shear and bulk viscosity
and heat conductivity) are sent to zero.
If the value of the transport coefficients were
known (e.g. by some first principle calculation),
then one could use dissipative hydrodynamics to
constrain e.g. the initial energy density distribution,
which is chosen conveniently in the ideal hydrodynamic
models. Or otherwise, choosing again 
physically acceptable initial conditions, one
is able to constrain the allowed ranges of the
transport coefficients. Despite recent progress
in first principle calculations \cite{Policastro:2001yc,Arnold:2003zc,Nakamura:2004sy,Arnold:2006fz,Huot:2006ys,Gagnon:2006hi,Aarts:2007wj,Meyer:2007ic,Buchel:2007mf,CaronHuot:2007gq},
the values of the hydrodynamic transport coefficients for
QCD in the relevant energy range are poorly 
constrained to date, so the second option
is currently the only viable possibility.

For RHIC, the first step in this direction was carried
out by Teaney \cite{Teaney:2003kp}, who provided estimates
for the sign and size of corrections due to
shear viscosity. This famous calculation,
however, did not provide a 
description of experimental data for non-zero viscosity,
because it was not dynamic and the initial conditions
could not be altered. Only very recently, the first
hydrodynamic calculations with shear viscosity describing
particle spectra for central and non-central collisions at RHIC
have became available 
\cite{Romatschke:2007jx,Romatschke:2007mq,Chaudhuri:2007qp}. 

Several other groups have produced numerical
codes capable of performing similar matching to data
\cite{Muronga:2004sf,Chaudhuri:2005ea,Muronga:2005pk,Chaudhuri:2006jd,Mota:2007tz,Song:2007fn,
Dusling:2007gi,Song:2007ux}.

However, the precise formulation of the viscous hydrodynamic
equations themselves has long been debated. To appreciate the complication,
one first has to understand a hydrodynamic formulation
for RHIC physics necessarily has to be fully relativistic,
and that the relativistic generalization of the Navier-Stokes
equations are acausal since they contain modes that
transport information at superluminal speeds.
These are high wavenumber modes and therefore
in principle outside the range of validity of
hydrodynamics, but in practice one has to find a way to
deal with them in viscous hydrodynamic simulations.
A possible solution to this problem is known as 
M\"uller-Israel-Stewart theory, where for each 
transport coefficient a corresponding relaxation time
is introduced which controls the speed of signal
propagation for the high wavenumber modes \cite{IS0a,IS0b,IS1,Mueller}.
For low momentum modes (up to first order in gradients),
the M\"uller-Israel-Stewart theory is identical to
the Navier-Stokes equations, but differs
for higher order gradients.
Unfortunately, this implied that 
the resulting equations retained
a certain degree of arbitrariness, as
it was not clear which additional 
terms of second or higher order in gradients
either within the M\"uller-Israel-Stewart or
other frameworks 
(see e.g.\cite{Muronga:2003ta,Heinz:2005bw,Baier:2006um,Koide:2006ef}) 
were allowed.
For the case of non-vanishing shear viscosity only,
it was shown recently \cite{Baier:2007ix} that the most general
form implies five independent terms of second order
in gradients. This form is general enough to describe
the hydrodynamic properties of (conformal) plasmas
both for weakly coupled systems
 describable by the Boltzmann
equation as well as infinitely strongly coupled
plasmas, which are accessible via Maldacena's conjecture
\cite{Maldacena:1997re}.

The aim of this work is to now apply this 
new set of equations for relativistic
shear viscous hydrodynamics to the problem of 
heavy-ion collisions at RHIC. 
In section \ref{sec:one}, we review the setup of conformal
relativistic viscous hydrodynamics and our numerics
for the simulation of heavy-ion collisions. In section \ref{sec:two},
details about the two main models of initial conditions
for hydrodynamics are given. Section \ref{sec:three} contains
our results for the multiplicity, radial flow and elliptic flow
in Au+Au collisions at top RHIC energies, as well as a note
on the notion of ``early thermalization''. We conclude in
section \ref{sec:four}.

\section{Setup}
\label{sec:one}

The energy-momentum tensor for relativistic
hydrodynamics in the presence of shear viscosity can be written as
\beq
T^{\mu \nu}=\epsilon u^\mu u^\nu-p \Delta^{\mu \nu}+\Pi^{\mu \nu},
\eeq
where $\epsilon$ and $p$ are the energy density and pressure, related 
by an equation of state $p=p(\epsilon)$. $u^\mu$ is the fluid 
four-velocity which fulfills $g_{\mu \nu} u^\mu u^\nu =1$, where the signature
of the metric is $g_{\mu \nu}=(+,-,-,-)$. The 
projector $\Delta^{\mu \nu}=g^{\mu \nu}-u^{\mu}u^\nu$ 
is orthogonal to the fluid velocity $u_\mu \Delta^{\mu \nu}=0$.
$\Pi^{\mu \nu}$ is the viscous shear tensor which is symmetric, traceless
($\Pi^\mu_\mu=0$) and orthogonal to the fluid velocity.
Hydrodynamics describes the evolution of the energy density and 
fluid velocity. The evolution equations are simply given by
the conservation of the energy momentum tensor
$D_\mu T^{\mu \alpha}=0$, 
where $D_\mu$ is the (geometric) covariant derivative.
Projection of $u_\alpha$ and $\Delta^\mu_\alpha$ on 
$D_\mu T^{\mu \alpha}=0$ gives 
\bqa
\label{hydro1}
(\epsilon+p)D u^\mu&=&\nabla^\mu p-
\Delta^\mu_\alpha D_\beta \Pi^{\alpha \beta}\, ,
\nonumber\\
D \epsilon &=& - (\epsilon+p) \nabla_\mu u^\mu+\frac{1}{2}\Pi^{\mu \nu}
\nabla_{\langle \nu} u_{\mu\rangle}\, ,
\eqa
where $D\equiv u^\alpha D_\alpha$ and $\nabla^\mu\equiv \Delta^{\mu \alpha} D_\alpha$ can be thought of as comoving
time and space derivatives, respectively. Note that 
$D_\mu = u_\mu D + \nabla_\mu$. The brackets $\langle\ \rangle$ denote
the combination
\beq
A_{\langle \mu} B_{\nu\rangle} =\left(\Delta^\alpha_\mu \Delta^\beta_\nu + 
\Delta^\alpha_\nu \Delta^\beta_\mu-\frac{2}{3} \Delta^{\alpha \beta} 
\Delta_{\mu \nu}\right) A_\alpha B_\beta,
\eeq
which is a projector that is symmetric, traceless, and orthogonal to the fluid velocity.
For later convenience, we also introduce symmetric and anti-symmetric
brackets
\bqa
A_{(\mu} B_{\nu)} &=&\frac{1}{2}\left(A_\mu B_\nu+A_\nu B_\mu\right)
\nonumber\\
A_{[\mu} B_{\nu]} &=&\frac{1}{2}\left(A_\mu B_\nu-A_\nu B_\mu\right).
\eqa

The equations (\ref{hydro1}) can be considered four equations
for the four independent components of $\epsilon, u^\mu$.
A theory of viscous hydrodynamics still has to specify 
the evolution or defining equations for the five independent components
of the shear tensor $\Pi^{\mu \nu}$. 
To first order in gradients, these are given by the 
relativistic Navier-Stokes equations
\beq
\Pi^{\mu \nu}=\eta \nabla^{\langle \nu} u^{\mu\rangle},
\label{NSeq}
\eeq
where $\eta$ is the shear viscosity coefficient.
As mentioned in the introduction, this theory suffers from
acausal signal propagation and associated numerical instabilities.
To second order in gradients, the evolution equations are given by
\cite{Baier:2007ix} (see also \cite{Bhattacharyya:2008jc})
\bqa
\Pi^{\mu\nu} &=& \eta \nabla^{\langle \mu} u^{\nu\rangle}
- \tau_\Pi \left[ \Delta^\mu_\alpha \Delta^\nu_\beta D\Pi^{\alpha\beta} 
 + \frac 4{3} \Pi^{\mu\nu}
    (\nabla_\alpha u^\alpha) \right] \nonumber\\
  &&\quad 
  + \frac{\kappa}{2}\left[R^{<\mu\nu>}+2 u_\alpha R^{\alpha<\mu\nu>\beta} 
      u_\beta\right]\nonumber\\
  && -\frac{\lambda_1}{2\eta^2} {\Pi^{<\mu}}_\lambda \Pi^{\nu>\lambda}
  +\frac{\lambda_2}{2\eta} {\Pi^{<\mu}}_\lambda \omega^{\nu>\lambda}
  - \frac{\lambda_3}{2} {\omega^{<\mu}}_\lambda \omega^{\nu>\lambda}\, ,
\label{maineq}
\eqa
where $\omega_{\mu \nu}=-\nabla_{[\mu} u_{\nu]}$ is the 
fluid vorticity and $R^{\alpha \mu \nu \beta},R^{\mu \nu}$
are the Riemann and Ricci tensors, respectively.
The coefficients $\tau_\Pi,\kappa,\lambda_1,\lambda_2,\lambda_3$
are the five new coefficients controlling the size of
the allowed terms of second order in gradients.
Having an application to the problem of heavy-ion collisions
in mind, the above set of equations can be simplified:
for all practical purposes spacetime can be considered flat, such
that both the Riemann and Ricci tensors vanish identically.
Thus, only the four coefficients 
$\tau_\Pi,\lambda_1,\lambda_2,\lambda_3$ enter the problem.

\subsection{A note on bulk viscosity and conformality}

Besides shear viscosity, QCD also has non-vanishing
bulk viscosity $\zeta$ which can be related to the 
QCD trace anomaly \cite{Kharzeev:2007wb}
\beq
\zeta\sim T^{\mu}_\mu=\epsilon-3 p.
\eeq
QCD lattice simulations seem to indicate that the ratio
bulk viscosity over entropy density $s$, $\zeta/s$,
is small compared to $\eta/s$ except for a small region
around the QCD deconfinement transition temperature, 
where it is sharply peaked \cite{Sakai:2007cm,Meyer:2007dy,Karsch:2007jc}.
If we are interested in describing effects
from shear viscosity only, we are led to consider $\zeta=0$, or
conformal fluids. This has been the main guiding principle
in Ref.~\cite{Baier:2007ix} and as a consequence Eq.~(\ref{maineq})
obeys conformal invariance, unlike most other second-order 
theories\footnote{Note that Muronga derived a version of 
Eq.~(\ref{maineq}) in Ref.~\cite{Muronga:2003ta} that 
turns out to obey conformal symmetry.}.

\subsection{First steps: 0+1 dimensions}
\label{section01}

In order to get a crude estimate of the 
effect of viscous corrections, let us consider the 
arguably simplest model of a heavy-ion collision: 
a system expanding in a boost-invariant fashion along
the longitudinal direction and having uniform energy
density in the transverse plane.
Introducing the Milne variables proper time 
$\tau=\sqrt{t^2-z^2}$ and space-time rapidity $\xi={\rm arctanh}(z/t)$,
boost invariance simply translates to requiring all
hydrodynamic variables ($\epsilon,u^\mu,\Pi^{\mu \nu}$)
to be independent of rapidity, and tensor components
$u^\xi,\Pi^{\mu \xi}$ to vanish. Assuming uniformity
in the transverse plane furthermore requires independence
from the transverse coordinates ${\bf x}_T=(x,y)$.
Even though this means that all the velocity components
except $u^\tau$ are zero,
the system is nevertheless non-trivial in the 
sense that the 
sum over velocity gradients
does not vanish, $\nabla_\mu u^\mu=\frac{1}{\tau}$,
sometimes referred to as ``Bjorken flow''.

In a way one has modeled an expanding system in static
space-time by a system at rest
in an expanding spacetime. This has been achieved by
transforming to the Milne coordinates $\tau,\xi$,
where the metric is 
$g_{\mu \nu}={\rm diag}(g_{\tau \tau},g_{xx},g_{yy},g_{\xi \xi})=
(1,-1,-1,-\tau^2)$.
Note that even though the spacetime in these coordinates
is expanding, it is nevertheless flat (e.g. has vanishing
Riemann tensor).

In this 0+1 dimensional toy model, the viscous hydrodynamic
equations become exceptionally simple \cite{Baier:2007ix},
\bqa
\partial_\tau \epsilon&=&-\frac{\epsilon+p}{\tau}+\frac{\Pi^\xi_\xi}{\tau}
\nonumber\\
\partial_\tau \Pi^\xi_\xi &=& -\frac{\Pi^\xi_\xi}{\tau_\Pi}
+\frac{4 \eta}{3 \tau_\Pi \tau}-\frac{4}{3 \tau} \Pi^\xi_\xi
-\frac{\lambda_1}{2\tau_\Pi\eta^2} \left(\Pi^\xi_\xi\right)^2.
\label{0+1dsystem}
\eqa
The Navier-Stokes equations are recovered formally in the limit
where all second-order coefficients vanish 
(e.g. $\tau_\Pi,\lambda_1\rightarrow 0$); then, one 
simply has 
\beq
\Pi^\xi_\xi=\frac{4 \eta}{3 \tau}.
\label{FOvalue}
\eeq
The equations (\ref{0+1dsystem}) can be solved numerically along the
lines of \cite{Muronga:2001zk,Baier:2006um}.
At very early times, where $\Pi^\xi_\xi>(\epsilon+p)$, 
the Navier-Stokes equations indicate an increase in energy density
and a negative effective longitudinal pressure $p-\Pi^\xi_\xi$.
Since gradients $\nabla_\mu u^\mu=1/\tau$ are strongest at early times,
this suggests that one is applying the Navier-Stokes equations
outside their regime of validity. Theories including second order
gradients may be better behaved at early times, but eventually
also have to break down when gradients become too strong. Here we want
to study the effects of the second order coefficients
on the value of the shear tensor at late times, where
a hydrodynamic approach should be valid.

To this end, let us study the deviation of the shear
tensor from its first order value, $\delta \Pi=\Pi^{\xi}_\xi-\frac{4 \eta}{3 \tau}$. At late times, Eq.~(\ref{0+1dsystem}) implies 
$\epsilon\sim \tau^{-4/3}$, so $\eta\sim \tau^{-1}$. Thus, if
$\delta \Pi$ is small compared to the first order value, from
Eq.~(\ref{0+1dsystem}) we find
\beq
\delta \Pi=\frac{4 \eta}{3 \tau}\left(\frac{2\tau_\Pi}{3\tau}
-\frac{2\lambda_1}{3 \tau \eta}\right).
\eeq
For a strongly coupled ${\cal N}=4$ plasma 
\cite{Policastro:2001yc,Baier:2007ix,Bhattacharyya:2008jc,Natsuume:2007ty}, 
one has\footnote{For completeness, we also mention the results
$\kappa=\frac{\eta}{\pi T},\lambda_2=-\frac{\eta \ln 2}{\pi T},\lambda_3=0$
from \cite{Baier:2007ix,Bhattacharyya:2008jc}.}
\beq
\frac{\eta}{s}=\frac{1}{4 \pi},\qquad
\tau_\Pi=\frac{2-\ln 2}{2 \pi T},\qquad 
\lambda_1=\frac{\eta}{2 \pi T},
\eeq
and thus $\Pi^\xi_\xi$ is larger than its first order value
by a factor of $1+\frac{1-\ln 2}{3 \pi T \tau}$.
For RHIC, $T \tau\gtrsim 1$ is a reasonable estimate, so one
finds that the second order corrections to $\Pi^\xi_\xi$ 
increase its value by a few percent over the first order result.

As an example on the importance of obeying conformal invariance, 
imagine dropping the term involving $\nabla_\alpha u^\alpha$ 
in the first line of Eq.~(\ref{maineq}). Redoing the above 
calculation one finds
\beq
\delta \Pi_{NC}=\frac{4 \eta}{3 \tau}\left(\frac{2\tau_\Pi}{\tau}
-\frac{2\lambda_1}{3 \tau \eta}\right),
\eeq
which indicates a nearly ten-fold increase of the size of $\delta \Pi$
for the non-conformal theory. For a weakly coupled plasma well described
by the Boltzmann equation \cite{Baier:2007ix}, where
one has $\tau_\Pi=\frac{6 \eta}{s T}$, ($\lambda_1$ is unknown but
generally set to zero in M\"uller-Israel-Stewart theory ),
the effect may be less pronounced, but still one qualitatively
expects second-order effects to be anomalously large if conformal
invariance is broken in an ``ad-hoc'' manner.

%
%

Clearly, the above estimates are not meant to be quantitative.
Indeed, even the sign of the correction may change when allowing
more complicated (e.g. three-dimensional) dynamics.
However, the lesson to be learned from this exercise is that
second-order gradients can and indeed do modify the 
shear tensor from its first order (Navier-Stokes) value.
This is physically acceptable, as long as the second-order
corrections are small compared to the first order ones (otherwise
the system is probably too far from equilibrium for even
a hydrodynamic description correct to second order in gradients 
to be valid). A practical means for testing this is 
calculating physical observables for different values of
the second-order coefficients and making sure that the
results do not strongly depend on the choice for these specific values.

\subsection{Including radial flow: lessons from 1+1 dimensions}

Some more insight on the effect of viscous corrections may be
gained by improving the model of the previous subsection
to allow for radially symmetric dynamics in the transverse plane
(but still assuming boost invariance).
This is most easily implemented by changing to polar coordinates
$(x,y)\rightarrow(r, \phi)$ with $r=\sqrt{x^2+y^2}$ and
$\phi={\rm arctan}(y/x)$. In this case, the only non-vanishing
velocity components are $u^\tau$ and $u^r$, and hence the
vorticity $\omega^{\mu \nu}$ vanishes identically.
Although non-trivial, the radially symmetric flow case is still 
a major simplification over the general form 
Eq.~(\ref{maineq}), since again the terms involving
$\kappa,\lambda_2,\lambda_3$ drop out.

Such a formulation allows both for important code
tests \cite{Baier:2006gy} as well as realistic simulations
of central heavy-ion collisions \cite{Romatschke:2007jx}
(note that truncated versions of Eq.~(\ref{maineq})
were used in these works).
The advantage of this formulation is that since the equations
are comparatively simple, it is rather straightforward to
implement them numerically and they are 
not very time consuming to solve since
only one dimensional (radial) dynamics is involved. 
The shortcoming of simulations with radially symmetric
flow profiles (``radial flow'') is that by construction
they cannot be matched to experimental data on the
impact-parameter dependence of multiplicity. Thus,
the considerable freedom in the initial/final conditions
inherent to all hydrodynamic approaches cannot be
eliminated in this case.

For this reason, we will choose not to discuss
the case of radial flow here in more detail, but
rather will comment on it later as a special
case of the more general situation.

\subsection{Elliptic flow: 2+1 dimensional dynamics}

Retaining the assumption of boost invariance, but
allowing for general dynamics in the transverse plane,
it is useful to keep Cartesian coordinates in the
transverse plane, and thus $u^\tau,u^x,u^y$
are the non-vanishing fluid velocities. The main reason
is that e.g. in polar coordinates
the equations
for the three independent components of $\Pi^{\mu \nu}$
would involve some extra non-vanishing Christoffel symbols
(other than $\Gamma^\tau_{\xi \xi}=\tau$,$\Gamma^\xi_{\tau \xi}=1/\tau$).

Fortunately, the case of two dimensions is special insofar
as the only nontrivial component of the vorticity tensor,
namely $\omega^{xy}$, fulfills the equation \cite{Romatschke:2007mq}
\beq
D \omega^{xy}+ \omega^{xy} \left[ \nabla_\mu u^\mu + \frac{D p}{\epsilon+p}
-\frac{D u^\tau}{u^\tau}\right]=\mathcal{O}(\Pi^3),
\label{vorteq}
\eeq
which can be derived by forming the combination 
$\nabla^x D u^y-\nabla^y D u^x$. The expression $\mathcal{O}(\Pi^3)$
denotes that the r.h.s. of Eq.~(\ref{vorteq}) is of third order in
gradients, and thus should be suppressed in the domain
of applicability of hydrodynamics.
For heavy-ion collisions, typically $\nabla_\mu u^\mu\ge\frac{1}{\tau}$,
so that for an equation of state with a speed of sound squared 
$c_s^2\equiv \frac{dp(\epsilon)}{d \epsilon}\sim \frac{1}{3}$,
Eq.~(\ref{vorteq}) translates to $\frac{D \omega^{xy}}{\omega^{xy}}<0$
unless $D \ln u^\tau\geq (1-c_s^2)\nabla_\mu u^\mu$.
In particular, this implies that in general 
$\omega^{xy}=0$ is a stable fix-point of the above equation and
hence we expect $\omega^{xy}$ to remain small throughout
the entire viscous hydrodynamic evolution if it is small initially.

Generically, one uses $u^{x,y}=0$ as an initial condition 
for hydrodynamics \cite{Huovinen:2006jp}, which implies $\omega^{xy}=0$
initially. Therefore, to very good approximation we can neglect the
terms involving vorticity in Eq.~(\ref{maineq}), such that
again only the second-order coefficients 
$\tau_\Pi,\lambda_1$ have to be specified.

The equations to be solved for 2+1 dimensional relativistic viscous hydrodynamics are
then (in components)
\bqa
(\epsilon+p) D u^i
&=&
c_s^2 \left(g^{ij}\partial_j \epsilon-u^i u^\alpha \partial_\alpha \epsilon
\right)-\Delta^i_\alpha D_\beta \Pi^{\alpha \beta}\nonumber\\
D \epsilon
&=&-(\epsilon+p) \nabla_\mu u^\mu+\frac{1}{2} \Pi^{\mu \nu}
\nabla_{\langle \mu} u_{\nu \rangle}\nonumber\\
D_\beta \Pi^{\alpha \beta}&=&
\Pi^{i \alpha}\partial_\tau \frac{u^i}{u^\tau}+
\frac{u^i}{u^\tau}\partial_\tau \Pi^{i \alpha}
+\partial_i \Pi^{\alpha i}
+\Gamma^{\alpha}_{\beta \delta} \Pi^{\beta \delta}
+\Gamma^\beta_{\beta \delta} \Pi^{\alpha \delta}\nonumber\\
\partial_\tau \Pi^{i \alpha}&=&
-\frac{4}{3 u^\tau}\Pi^{i \alpha} \nabla_\beta u^\beta
+\frac{\eta}{\tau_\Pi u^\tau}\nabla^{\langle i} u^{\alpha \rangle}
-\frac{1}{\tau_\Pi u^\tau}\Pi^{i \alpha}
\nonumber\\
&&-\frac{u^i \Pi^{\alpha}_\kappa + u^\alpha \Pi^{i}_\kappa}{u^\tau}
Du^\kappa-\frac{u^j}{u^\tau} \partial_j \Pi^{i \alpha}
-\frac{\lambda_1}{2 \eta^2 \tau_\Pi u^\tau}
\Pi^{\langle i}_\lambda \Pi^{\alpha \rangle \lambda}\nonumber\\
\nabla_\mu u^\mu &=&\partial_\tau u^\tau+\partial_i u^i
+\frac{u^\tau}{\tau}\nonumber\\
\nabla_{\langle x} u_{x \rangle}&=&
2 \Delta^{\tau x}\partial_\tau u^x
+2 \Delta^{i x}\partial_i u^x-\frac{2}{3}\Delta^{xx} \nabla_\mu u^\mu
\nonumber\\
\nabla_{\langle x} u_{y \rangle}&=&
\Delta^{\tau x}\partial_\tau u^y
+\Delta^{\tau y}\partial_\tau u^x
+\Delta^{i x}\partial_i u^y
+\Delta^{i y}\partial_i u^x-\frac{2}{3}\Delta^{xy} \nabla_\mu u^\mu\nonumber\\
\nabla_{\langle \xi} u_{\xi \rangle}&=&
2 \tau^4 \Delta^{\xi \xi}\Gamma^\xi_{\tau \xi} u^\tau
-\frac{2}{3}\tau^4\Delta^{\xi\xi} \nabla_\mu u^\mu.
\label{manyeq}
\eqa
Here and in the following 
Latin indices collectively denote the transverse coordinates $x,y$ and
the relation $u_\mu \Pi^{\mu \nu}=0$ has been used to derive 
the above equations (similarly, 
$u^\mu \nabla_{\langle \mu} u_{\nu \rangle}=0$ can be used to obtain 
the other non-trivial components needed).
Note that this particular form of Eq.~(\ref{manyeq}) 
has not been simplified further since it 
roughly corresponds to the equations 
implemented for the numerics of \cite{Romatschke:2007mq},
and is meant to facilitate understanding of the code \cite{codedown}.
A simple algorithm to solve Eq.~(\ref{manyeq}) has been
outlined in \cite{Baier:2006gy} and will be reviewed in the next 
subsection for completeness.

\subsection{A numerical algorithm to solve relativistic viscous hydrodynamics}

The first step of the algorithm consists of choosing the 
independent degrees of freedom. For boost-invariant 2+1 dimensional dynamics,
a sensible choice for this set 
is e.g. $\epsilon$, $u^x$, $u^y$, $\Pi^{xx}$, $\Pi^{xy}$, $\Pi^{yy}$.
The pressure is then obtained via the equation of state $p(\epsilon)$,
and the only other non-vanishing velocity as $u^\tau=\sqrt{1+u^2_x+u^2_y}$.
Similarly, the other nonzero components of $\Pi^{\mu \nu}$ are calculated
using the equations $\Pi^{\mu}_\mu=0$, $u_\mu \Pi^{\mu \nu}=0$.

Given the value of the set of independent components at some
time $\tau=\tau_0$, the aim is then to construct an algorithm
from Eq.~(\ref{manyeq}) such that the new values of the set can be
calculated as time progresses. Note that in Eq.~(\ref{manyeq}), time derivatives
of the independent component set enter only linearly.
Therefore, Eq.~(\ref{manyeq}) may be written as a matrix equation
for the derivatives of the independent component set,
\beq
\left(
\begin{array}{cccc}
a_{00} & a_{01} & \ldots  &a_{05}\\
a_{10} & a_{11} & \ldots  &a_{15}\\
\multicolumn{4}{c}\dotfill\\
a_{50} & a_{51} & \ldots  &a_{55}
\end{array}
\right)
\cdot
\left(\begin{array}{c}
\partial_\tau \epsilon\\
\partial_\tau u^x\\
\ldots\\
\partial_\tau \Pi^{yy}
\end{array}\right)
=\left(\begin{array}{c}
b_0\\
b_1\\
\ldots\\
b_6
\end{array}\right).
\eeq
Denoting the above matrix and vector as ${\bf A}$ and ${\bf b}$, respectively,
a straightforward way to obtain the time derivatives is via 
numerical matrix inversion,
\beq
\left(\begin{array}{c}
\partial_\tau \epsilon\\
\partial_\tau u^x\\
\ldots\\
\partial_\tau \Pi^{yy}
\end{array}\right) = {\bf A}^{-1} \cdot {\bf b}.
\label{simpleeq}
\eeq
Choosing a naive discretization of derivatives
\beq
\partial_\tau f(\tau)=\frac{f(\tau+\delta \tau)-f(\tau)}{\delta \tau},
\qquad
\partial_x f(x)=\frac{f(x+a)-f(x-a)}{2a},
\eeq
which is first order accurate in the temporal grid spacing $\delta \tau$ 
and second order accurate in the spatial grid spacing $a$,
one can then directly calculate the new values of the independent
component set from Eq.~(\ref{simpleeq}). 

Note that for ideal hydrodynamics, the algorithm Eq.~(\ref{simpleeq}) 
would fail for this naive discretization \cite{NR}.
The reason is that ideal hydrodynamics is inherently unstable
to high wavenumber fluctuations (which can be thought of as the basis
for turbulence). For ideal hydrodynamics, one thus has to
use a discretization which amounts to the introduction of
numerical viscosity to dampen these fluctuations.
Luckily, viscous hydrodynamics does not suffer from this
problem because it has real, physical viscosity inbuilt.
It is because of this reason that the naive discretization
can be used in the algorithm Eq.~(\ref{simpleeq})
without encountering the same problems as in ideal hydrodynamics,
as long as a finite value for the viscosity $\eta$ is used\footnote{
In practice we have used $\frac{\eta}{s}>10^{-4}$. Typically,
between $\frac{\eta}{s}=10^{-2}$ and $\frac{\eta}{s}=10^{-4}$ there 
are no significant changes in the hydrodynamic results and we 
refer to $\frac{\eta}{s}=10^{-4}$ as ``ideal hydrodynamics''.}.
While applicable to sufficiently smooth initial conditions,
the above algorithm is too simple to treat strong gradients
such as the propagation of shocks, and should be replaced
by a more involved scheme in such cases.

Since matrix inversions are computationally intensive,
one can speed up the numerics by expressing 
$\partial_\tau \Pi^{\mu \nu}$ in terms of 
$\partial_\tau u^i,\partial_\tau \epsilon$. Inserting these
in the equations for $D u^\mu$ and $D \epsilon$,
one only needs to invert a $3\times3$ matrix to obtain
the new values of the energy density and fluid velocities.
This approach has been used in 
\cite{Romatschke:2007jx,Romatschke:2007mq,Baier:2006gy}.

\subsection{Initial conditions and equation of state}

As outlined in the introduction, any hydrodynamic
description of RHIC physics relies on given initial 
energy density distributions. Two main classes of models
for boost-invariant setups exist: the Glauber models
and the Color-Glass-Condensate (CGC) models.

As will be shown in the following, both model classes can give a reasonable
description of the experimentally found multiplicity distribution,
but they differ by their initial spatial eccentricity.
A detailed discussion of the initial conditions will be given
in subsequent sections.

Besides an initial condition for the energy density,
one also needs to specify an initial condition for
the independent components of the fluid velocities and the
shear tensor. For the fluid velocities we will follow
the standard assumption that these vanish initially
\cite{Huovinen:2006jp}. Finally, when using the set of equations
(\ref{manyeq}), one also has to provide initial values for
the independent components of $\Pi^{\mu \nu}$. Extreme choices
are $\Pi^{\mu \nu}=0$ and a shear tensor so large that
a diagonal component of the energy-momentum tensor vanishes
in the local rest frame (e.g. $\Pi^\xi_\xi=p$, or zero longitudinal
effective pressure),
with the physical result expected somewhere in between (see e.g.
the discussion in \cite{Dumitru:2007qr}).

Once the initial conditions for the independent hydrodynamic
variables have been specified, one needs the equation of
state to solve the hydrodynamic equations (\ref{manyeq}).
Aiming for a description of deconfined nuclear matter at zero chemical potential,
a semi-realistic equation of state has to incorporate
evidence from lattice QCD calculations \cite{Aoki:2006we} that the transition
from hadronic to deconfined quark matter is probably 
an analytic crossover, not a first or second order phase transition
as often used in ideal hydrodynamic simulations.
On the other hand, continuum extrapolations for the value of 
the energy density and pressure
for physical quark masses are still not accessible with high precision
using current lattice methods. For this reason, we will employ
the equation of state by Laine and Schr\"oder \cite{Laine:2006cp},
which is derived from a hadron resonance gas at low temperatures,
a high-order weak-coupling perturbative QCD calculation at high temperatures,
and an analytic crossover regime interpolating between the high
and low temperature regime, respectively.
\begin{figure}[t]
\center
\includegraphics[width=.5\linewidth]{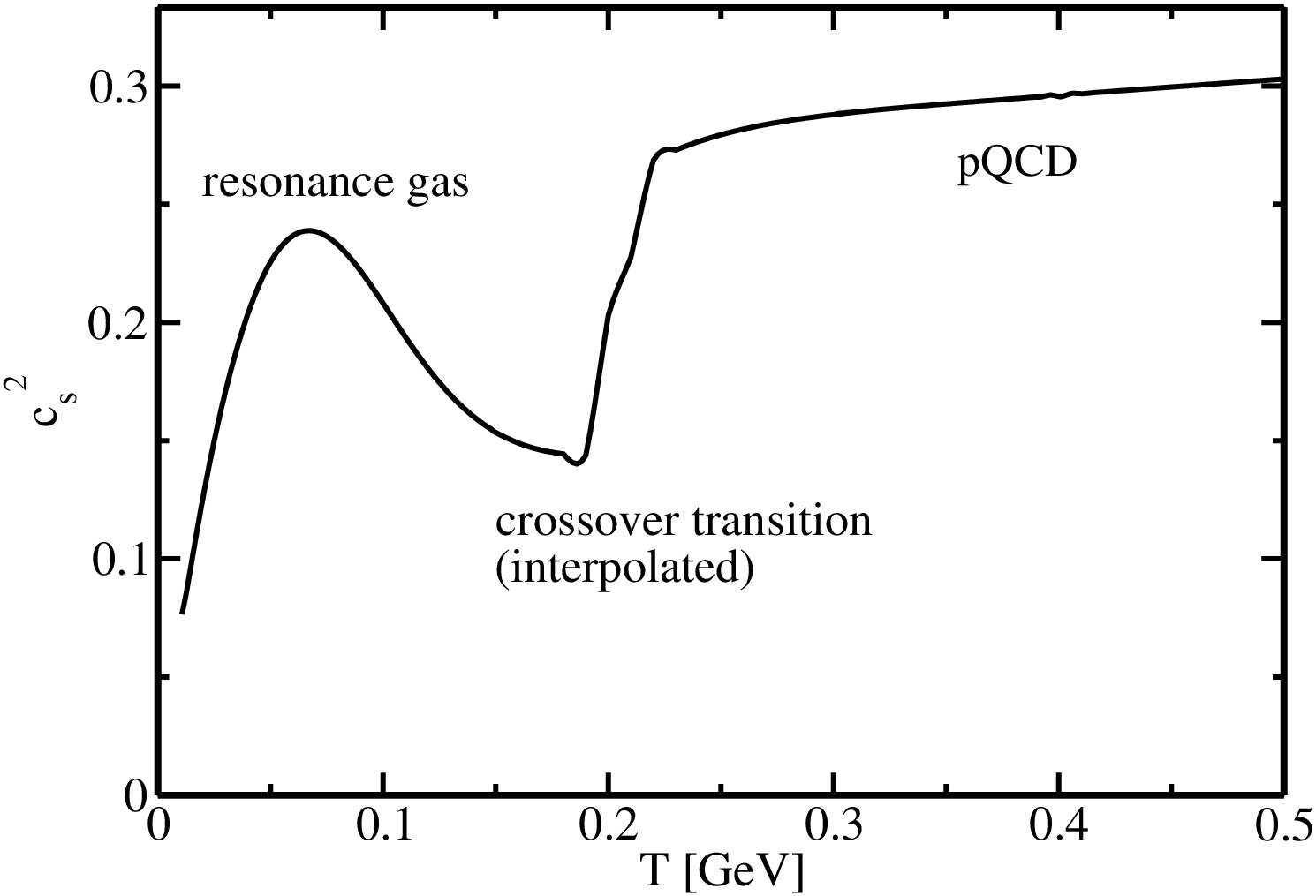}
\caption{The speed of sound squared from Ref.~\cite{Laine:2006cp},
used in the hydrodynamic simulations. See text for details.}
\label{figcs2}
\end{figure}
For hydrodynamics, an important quantity is the speed of sound
squared extracted from the equation of state, 
$c_s^2\equiv \frac{d p(\epsilon)}{d \epsilon}$. For completeness,
we reproduce a plot of this quantity in Fig.~\ref{figcs2}.

\subsection{Freeze-out}
\label{sec:fo}

At some stage in the evolution of the matter produced in 
a heavy-ion collision, the system will become too dilute
for a hydrodynamic description to be applicable.
This ``freeze-out'' process is most probably happening 
gradually, but difficult to model realistically.
A widely-used approximation is therefore to assume
instantaneous freeze-out whenever a certain fluid cell
cools below a certain predefined temperature or energy
density (see \cite{Hung:1997du,Dusling:2007gi} for different approaches). 
The standard prescription for this freeze-out
process is the Cooper-Frye formula \cite{CooperFrye}, which allows
conversion of the hydrodynamic variables (energy density, 
fluid velocity,...) into particle distributions.

Specifically, in the case of isothermal freeze-out at a
temperature $T_f$, the conversion from hydrodynamic
to particle degrees of freedom will have to take place
on a three-dimensional freeze-out hypersurface $\Sigma$,
which can be characterized by its normal four-vector, and parametrized by three space-time 
variables \cite{Ruuskanen:1986py,Rischke:1996em}. The spectrum
for a single particle on mass shell with four momentum 
$p^\mu=(E,{\bf p})$ and degeneracy $d$ is then given by
\beq
E\frac{d^3 N}{d^3 {\bf p}}\equiv 
\frac{d}{(2 \pi)^3} \int p_\mu d\Sigma^\mu f\left(x^\mu,p^\mu\right),
\label{CF}
\eeq
where $d\Sigma^\mu$ is the normal vector on the hypersurface 
$\Sigma$ and $f$ is the off-equilibrium distribution function.

Originally, the Cooper-Frye prescription was derived for
systems in thermal equilibrium, where $f$ is
built out of a Bose or Fermi distribution function 
$f_0(x)=\left(\exp[(x)\pm 1]^{-1}\right)$,
depending on the statistics of the particle under consideration.
In order to generalize it to systems out of equilibrium,
one customarily relies on the ansatz used in the derivation
of viscous hydrodynamics from kinetic theory 
\cite{deGroot},
\beq
f\left(x^\mu,p^\mu\right)=f_0\left(\frac{p_\mu u^\mu}{T}\right)
+f_0\left(\frac{p_\mu u^\mu}{T}\right) \left[1\mp 
f_0\left(\frac{p_\mu u^\mu}{T}\right)\right] \frac{p_\mu p_\nu \Pi^{\mu \nu}}
{2 T^2 (\epsilon+p)}.
\label{fullfansatz}
\eeq
For simplicity, in the following we approximate 
$f_0(x)\sim \exp(-x)$, so similarly
\beq
f\left(x^\mu,p^\mu\right)=\exp\left(-p_\mu u^\mu/T\right)
\left[1+
\frac{p_\mu p_\nu \Pi^{\mu \nu}}
{2 T^2 (\epsilon+p)}\right].
\eeq
The effect of this approximation will be commented on 
in the following sections.

In practice, for boost-invariant 2+1 dimensional hydrodynamics,
the freeze-out hypersurface 
$\Sigma^\mu=\left(\Sigma^t,\Sigma^x,\Sigma^y,\Sigma^z\right)=(t,x,y,z)$
can be parametrized either
by $\tau,\xi$ and the polar angle $\phi$, or by $x,y,\xi$:
\beq
\begin{array}{c}
t=\tau \cosh \xi\\
x=x(\tau, \phi)\\
y=y(\tau, \phi)\\
z=\tau \sinh \xi
\end{array}\quad,\qquad \qquad
\begin{array}{c}
t=\tau(x,y) \cosh \xi\\
x=x\\
y=y\\
z=\tau(x,y) \sinh \xi
\end{array}.
\eeq
The normal vector on $\Sigma^\mu$ is calculated by
\bqa
d\Sigma_\mu(\tau,\phi,\xi) &=& \varepsilon_{\mu \alpha \beta \gamma}
\frac{\partial \Sigma^\alpha}{\partial \tau}
\frac{\partial \Sigma^\beta}{\partial \phi}
\frac{\partial \Sigma^\gamma}{\partial \xi}
d\tau d\phi d\xi\nonumber\\
d\Sigma^\mu(\tau,\phi,\xi)&=&-\tau \left(\cosh \xi 
\left(\frac{\partial x}{\partial \tau}\frac{\partial y}{\partial \phi}-
\frac{\partial y}{\partial \tau}\frac{\partial x}{\partial \phi}\right),
\frac{\partial y}{\partial \phi},-\frac{\partial x}{\partial \phi},
\sinh \xi 
\left(\frac{\partial x}{\partial \tau}\frac{\partial y}{\partial \phi}-
\frac{\partial y}{\partial \tau}\frac{\partial x}{\partial \phi}\right)
\right) d\tau d\phi d\xi \nonumber
\eqa
and similarly for the other 
parametrization \cite{Kolb:2003dz}.

\begin{figure}[t]
\center
\includegraphics[width=.5\linewidth]{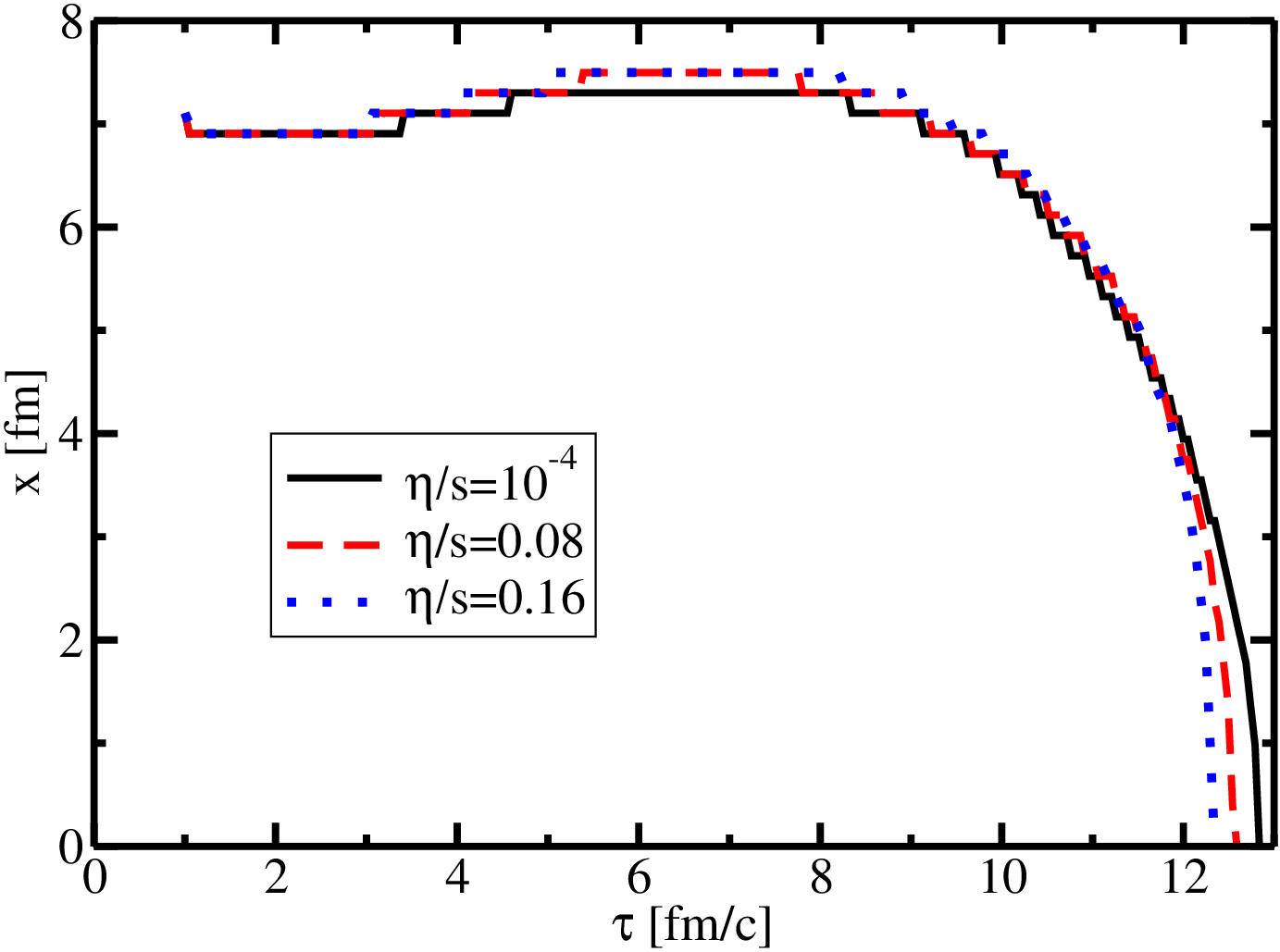}
\caption{(Color online) Space-time cut through the three-dimensional hypersurface
for a central collision within the Glauber model. Simulation parameters
used were $a=1$ GeV$^{-1}$, $\tau_0=1$ fm/c, $T_i=0.36$ GeV, 
$T_f=0.15$ GeV, $\tau_\Pi=6 \frac{\eta}{s}$ and $\lambda_1=0$ (see
next sections for definitions). As can be seen from the figure,
inclusion of viscosity only slightly changes the form of the surface.}
\label{freeze}
\end{figure}

For a realistic equation of state, at early times
the freeze-out hypersurface will contain the same transverse coordinate values 
$(x,y)$ for different times $\tau$ (see Fig.~\ref{freeze}). Therefore, the 
parametrization in terms of $(x,y,\xi)$ cannot be used for early times. 
On the other hand,
the parametrization in terms of $(\tau,\phi,\xi)$ 
contains derivatives of $(x,y)$
with respect to $\tau$, which become very large at late times (see
Fig.~\ref{freeze}).
Numerically, it is therefore not advisable to use this parametrization
at late times.
As a consequence, we use the one parametrization at
early times but switch to the other parametrization at late times,
such that the integral in Eq.~(\ref{CF}) is always defined and
numerically well-behaved\footnote{It may be possible that other
parametrizations may turn out to be more convenient. For instance,
it is conceivable that performing a triangulation of the three-dimensional
hypersurface and replacing the integral in (\ref{CF}) by a sum
over triangles could turn out to be numerically superior to our method.}.

In order to evaluate the integral (\ref{CF}), it is useful
to express $p^\mu$ also in Milne coordinates, 
\beq
p^\mu=(p^\tau,p^x,p^y,p^\xi)=(m_T \cosh (Y-\xi),p^x,p^y,  
\frac{m_T}{\tau}\sinh (Y-\xi)),
\eeq
where $m_T=\sqrt{m^2+p_x^2+p_y^2}=\sqrt{E^2-p_z^2}$. Here and
in the following $Y={\rm arctanh}(p^z/E)$ is the rapidity, 
and $m$ is the rest mass of the
particle under consideration.
Then the $\xi$ integration can be carried out analytically using
\beq
\frac{1}{2}
\int_{-\infty}^\infty d\xi \cosh^n(Y-\xi) \exp[-z \cosh(Y-\xi)]=
(-1)^n \partial_z^n K_0(z) \equiv K(n,z),
\eeq
where $K_0(z)$ is a modified Bessel function.
One finds
\bqa
E \frac{d^3 N}{d^3{\bf p}}&=&
\frac{2 d}{(2 \pi)^3} \int d\tau d\phi \exp{[(p^x u^x+p^y u^y)/T]}\times
\nonumber\\
&&\left[m_T 
\left(\frac{\partial x}{\partial \tau}\frac{\partial y}{\partial \phi}-
\frac{\partial y}{\partial \tau}\frac{\partial x}{\partial \phi}\right)
\left(T_1 K(1,m_T u^\tau/T)+T_2  K(2,m_T u^\tau/T)+T_3 K(3,m_T u^\tau/T)
\right)\right.\nonumber\\
&&\left.-\left(p^x \frac{\partial y}{\partial \phi}-
p^y \frac{\partial x}{\partial \phi}\right)
\left(T_1 K(0,m_T u^\tau/T)+T_2  K(1,m_T u^\tau/T)+T_3 K(2,m_T u^\tau/T)
\right)\right]\ ,\nonumber\\
T_1&=&1+\frac{m_T^2 \Pi^\xi_\xi+p_x^2 \Pi^{xx}+p_y^2 \Pi^{yy}+2 p_x p_y
\Pi^{xy}}{2 T^2 (\epsilon+p)}\ ,\nonumber\\
T_2&=&-2 m_T  \frac{p^x \Pi^{x \tau}+p^y \Pi^{y \tau}}{2 T^2 (\epsilon+p)}\ ,
\nonumber\\
T_3&=&m_T^2\frac{\Pi^{\tau \tau}-\Pi^\xi_\xi}{2 T^2 (\epsilon+p)}\ ,
\eqa
for the $(\tau,\phi,\xi)$ parametrization,
and a similar result for the other parametrization of the hypersurface.
The remaining integrals for the particle spectrum have 
to be carried out numerically unless one is considering
the case of a central collision 
\cite{Baier:2006gy,Romatschke:2007jx} where the 
integral has an additional symmetry in $\phi$.

For the simulation of a heavy-ion collision,
one then also needs to take into account the 
feed-down process of particle resonances that
decay into lighter, stable particles 
\cite{Sollfrank:1990qz,Sollfrank:1991xm}. 
Therefore, we calculate the spectra for particle
resonances with masses up to $\sim 2$ GeV and
then use available routines from
the AZHYDRO package \cite{OSCAR} to determine the spectra
of stable particles including these feed-down
contributions. Ultimately, one would be interested in
describing the last stage of the evolution by coupling the hydrodynamics 
to a hadronic cascade code \cite{Bass:2000ib,Teaney:2001av,Hirano:2005xf,Nonaka:2006yn}.
We leave this for future work.

The particle spectra $E\frac{d N_{\rm corr}}{d^3 {\bf p}}$ including
feed-down contributions
can then be used to calculate experimental observables at central
rapidity $Y=0$ ,
such as radial and elliptic flow coefficients, $v_0,v_2$, respectively.
These are defined as
\beq
v_0(p_T,b)=\int \frac{d \phi_p}{2 \pi} E\frac{d N_{\rm corr}}{d^3 {\bf p}}\ ,
\qquad
E\frac{d N_{\rm corr}}{d^3 {\bf p}}=v_0(p_T,b) \left[1+2 v_2 (p_T,b) \cos (2 \phi_p)
+\ldots\right],
\eeq
where $\phi_p=\arctan (p^y/p^x)$ and $p_T=\sqrt{p_x^2+p_y^2}$. Furthermore, 
the total multiplicity per unit rapidity $\frac{dN}{dy}$ and 
the mean transverse momentum $<p_T>$ 
are then given by
\beq
\frac{d N}{dy}\equiv 2\pi \int dp_T\ p_T\ v_0(p_T,b)\ ,
\qquad
<p_T>\equiv \frac{\int d p_T\  p_T^2 \ v_0(p_T,b)}
{\int d p_T\ p_T\ v_0 (p_T,b)}.
\eeq
The $p_T$ integrated elliptic flow coefficient is defined as
\beq
v_2^{\rm int}(b)=\frac{\int d p_T\ p_T v_2(p_T,b) v_0(p_T,b)}
{\int d p_T\ p_T\ v_0 (p_T,b)}
\eeq
and the minimum bias elliptic flow coefficient as \cite{Kolb:2001qz}
\beq
v_2^{{\rm mb}}(p_T)=\frac{\int db\ b\ v_2(p_T,b)\ v_0(p_T,b)}
{\int db\ b\ v_0 (p_T,b)}.
\eeq

\subsection{Code tests}

It is imperative to subject the numerical implementation
of the relativistic viscous hydrodynamic model to several tests.
The minimal requirement is that the code is stable for a range
of simulated volumes and grid spacings $a$, such that 
an extrapolation to the continuum may be attempted (keeping the 
simulated volume fixed but sending $a\rightarrow 0$).
Our code fulfills this property.

Furthermore, one has to test whether 
this continuum extrapolation corresponds to the correct physical
result in simple test cases.
One such test case is provided by the 0+1 dimensional
model discussed in section \ref{section01}. Using initial
conditions of uniform energy density in the 2+1 dimensional
numerical code, the temperature evolution should match
that of Eq.~(\ref{0+1dsystem}), for which it is straightforward 
to write an independent numerical solver. 
Our 2+1 dimensional code passes this test, for small and large
$\eta/s$ and different values for $\tau_\Pi,\lambda_1$.

The above test is non-trivial in the sense that it
allows to check the implementation of
nonlinearities in the hydrodynamic model.
However, it does not probe the dynamics of the model,
since e.g. all velocities are vanishing.
Therefore, another test that one can (and should!) conduct is to
study the dynamics of the model against that
of linearized hydrodynamics (this test was first
outlined in Ref.~\cite{Baier:2006gy}; see \cite{Bhalerao:2007ek} for similar
considerations). More specifically,
let us consider a viscous background ``solution'' with 
$u^i=0$ but non-vanishing $\epsilon(\tau),\Pi^\xi_\xi(\tau)$
obeying Eq.~(\ref{0+1dsystem}). To first order in small fluctuations 
$\delta \epsilon,\delta u^\mu, \delta \Pi^{\mu \nu}$
around this background the set of equations (\ref{manyeq}) become
\bqa
\left[c_s^2 \partial_\tau \epsilon+\frac{1}{2}\partial_\tau \Pi^\xi_\xi
+\frac{3}{2 \tau} \Pi^\xi_\xi
+(\epsilon+p+\frac{1}{2}\Pi^\xi_\xi)\partial_\tau \right] \delta u^x
+c_s^2 \partial_x \delta \epsilon + \partial_i \delta \Pi^{x i}
&=&0\nonumber\\
\left[c_s^2 \partial_\tau \epsilon+\frac{1}{2}\partial_\tau \Pi^\xi_\xi
+\frac{3}{2 \tau} \Pi^\xi_\xi
+(\epsilon+p+\frac{1}{2}\Pi^\xi_\xi)\partial_\tau \right] \delta u^y
+c_s^2 \partial_y \delta \epsilon + \partial_i \delta \Pi^{y i}
&=&0\nonumber\\
\left[\partial_\tau +\frac{1+c_s^2}{\tau} \right]\delta\epsilon+\left[(\epsilon+p)+\frac{1}{2}\Pi^\xi_\xi\right] \partial_i \delta u^i
-\frac{1}{\tau}\delta \Pi^\xi_\xi
&=&0\nonumber\\
\left[\frac{4}{3\tau}+\frac{1}{\tau_\Pi}+\partial_\tau\right]
\delta \Pi^\xi_\xi
-\left[\frac{4 \eta}{3 \tau \tau_\Pi}+ \frac{1}{4\tau_\Pi} \Pi^\xi_\xi\right] \frac{\delta \epsilon}{\epsilon}
+\left[\frac{2 \eta}{3 \tau_\Pi} +\frac{4}{3} \Pi^\xi_\xi\right]
\partial_i \delta u^i&=&0\nonumber\\
\left[\frac{4}{3\tau}+\frac{1}{\tau_\Pi}+\partial_\tau\right]
\delta \Pi^{xx}
-\left[\frac{2 \eta}{3 \tau_\Pi \tau}+ \frac{1}{4\tau_\Pi} \Pi^{xx}\right]\frac{\delta \epsilon}{\epsilon}
+\frac{2 \eta}{\tau_\Pi}\partial_x \delta u^x
+\left[-\frac{2 \eta }{3 \tau_\Pi}+\frac{4}{3} \Pi^{xx}\right]
\partial_i \delta u^i&=&0\nonumber\\
\left[\frac{4}{3\tau}+\frac{1}{\tau_\Pi}+\partial_\tau\right]
\delta \Pi^{xy}
+\frac{\eta}{\tau_\Pi}\left(\partial_x \delta u^y+\partial_y
\delta u^x\right)&=&0,\qquad \
\label{linearizedeq}
\eqa
where we have put $\lambda_1=0$ and assumed a constant $c_s^2$ for simplicity. 
Noting that $\delta \Pi^{yy}=\delta \Pi^\xi_\xi-\delta \Pi^{xx}$
from $\delta \Pi^\mu_\mu=0$,  Eq.~(\ref{linearizedeq}) 
are a closed set of linear, but coupled differential equations
for the fluctuations 
$\delta \epsilon,\delta u^x,\delta u^y,\delta \Pi^\xi_\xi, \delta \Pi^{xx},\delta \Pi^{xy}$. Doing a Fourier transform,
\beq
\delta \epsilon(\tau,x,y)=\int \frac{d^2 {\bf k}}{(2 \pi)^2}
e^{i x k^x+i y k^y} \delta \epsilon(\tau,k^x, k^y)
\eeq
(and likewise for the other fluctuations), Eq.~(\ref{linearizedeq})
constitute coupled ordinary differential equations for each
mode doublet ${\bf k}=(k^x,k^y)$, which again are straightforward
to solve with standard numerical methods \cite{codedown} (and analytically for 
ideal hydrodynamics).

A useful test observable is the correlation function
\beq
f(\tau,{\bf x_1},{\bf x_2})=\frac{\left<\delta \epsilon(\tau,{\bf x_1}) 
\delta \epsilon(\tau,{\bf x_2})\right>}{\epsilon(\tau)^2},
\eeq
where $\left< \right>$ denotes an ensemble average
over initial conditions $\left.\delta \epsilon\right|_{\tau=\tau_0}$.
In particular, let us study initial conditions where $\delta \epsilon$
is given by Gaussian random noise with standard deviation $\Delta$, 
\beq
f(\tau_0,{\bf x_1},{\bf x_2})=\Delta^2 \delta^2({\bf x_1-x_2})
\eeq
and all other fluctuations vanish initially. These initial
conditions are readily implemented both for the full 2+1 dimensional
hydrodynamic code as well as for the linearized system 
Eq.~(\ref{linearizedeq}). As the system evolves to finite time $\tau$,
both approaches have to give the same correlation function $f$ as 
long as the linearized treatment is applicable, and hence
Eq.~(\ref{linearizedeq}) can be used to test the dynamics of 
the full numerical code.

\begin{figure}[t]
\center
\includegraphics[width=.5\linewidth]{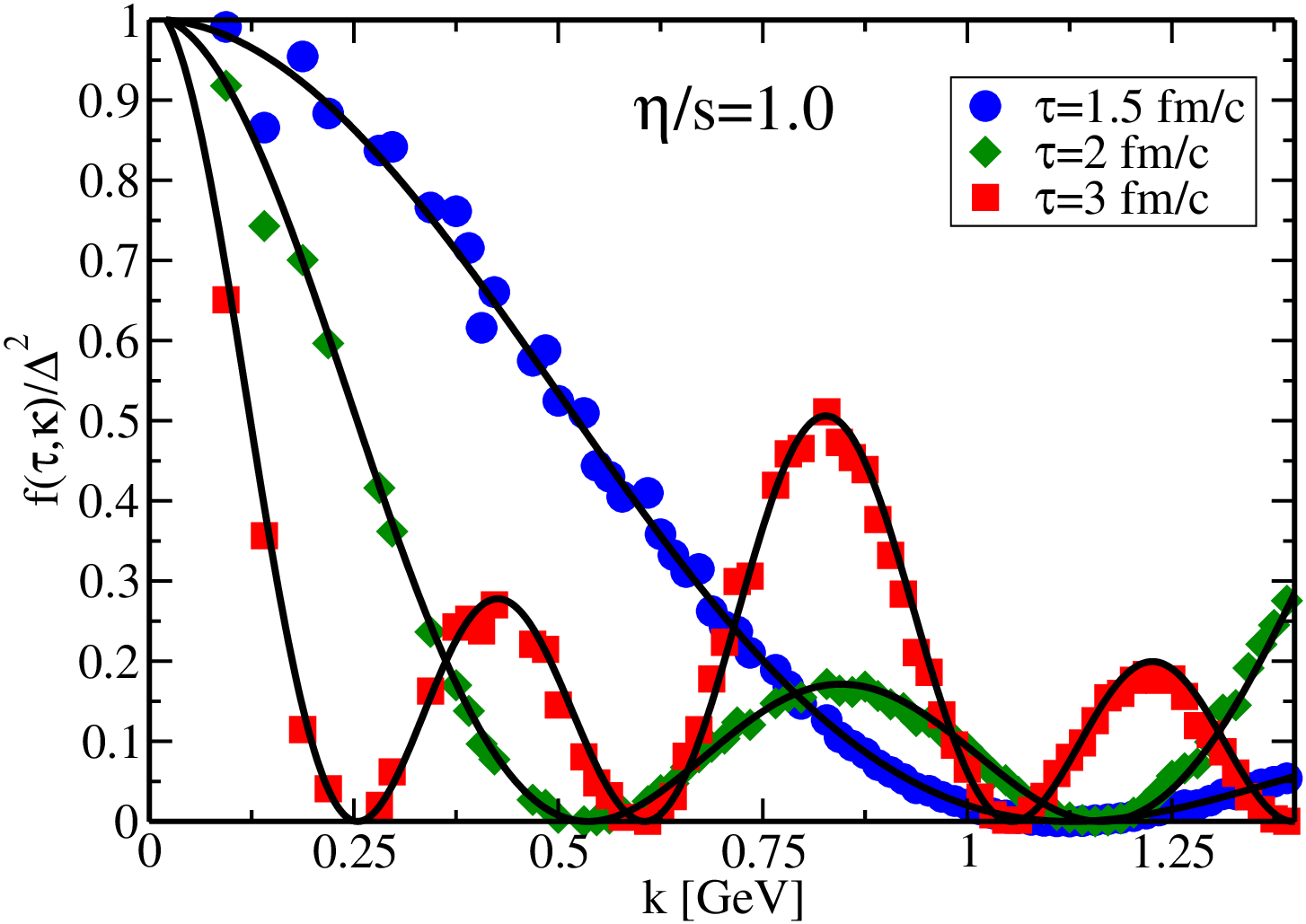}
\caption{(Color online) The correlation function $f(\tau,{\bf k})$ 
as a function of momentum $k=|{\bf k}|$
for a lattice with $a=1$ GeV$^{-1}$, $64^2$ sites and averaged over
30 initial configurations (symbols),
compared to the result from the linearized hydrodynamic
equations (lines).}
\label{figflucs}
\end{figure}

In practice, note that for the above construction $f$ can only
depend on the difference of coordinates,
\beq
\frac{\left<\delta \epsilon(\tau,{\bf x_1}) 
\delta \epsilon(\tau,{\bf x_2})\right>}{\epsilon(\tau)^2}
=f(\tau,{\bf x_1}-{\bf x_2})=\int \frac{d^2{\bf k}}{(2\pi)^2} e^{i {\bf k}\cdot
({\bf x_1-x_2})} f(\tau,{\bf k})
\eeq
and therefore in Fourier-space
\beq
 f(\tau,{\bf k}) \delta^2({\bf k^\prime}) =
\frac{\left<\delta \epsilon(\tau,{\bf k}) 
\delta \epsilon(\tau,{\bf k^\prime-k})\right>}{(2 \pi)^2 \epsilon(\tau)^2}.
\label{corrdef}
\eeq
In the full 2+1 dimensional numerical code which is discretized
on a space-time lattice, $\delta^2({\bf k^\prime})$
is regular for any finite $a$, and one can maximize the signal
for $f(\tau,{\bf k})$ by calculating the r.h.s. 
of Eq.~(\ref{corrdef}) for ${\bf k^\prime}=0$. 
Similarly, one solution $\delta \epsilon (\tau,{\bf k})$ per ${\bf k}$ mode
is sufficient calculate $f(\tau,{\bf k})$
for the linearized system Eq.~(\ref{linearizedeq}).

The above initial conditions imply $f(\tau=\tau_0,{\bf k})=\Delta^2$,
but for finite times characteristic peaks develop as a function
of $|{\bf k}|$, whose position, height
and width are sensitive to the values of $c_s^2,\tau_\Pi,\eta/s$ and
of course the correct implementation of the hydrodynamic equations.
The comparison between full numerics and linearized treatment
shown in Fig.~\ref{figflucs} suggests that our code also passes
this test\footnote{Note that a small numerical
error occurred in the linearized hydrodynamic solver 
and the corresponding figure in Ref.~\cite{Romatschke:2007mq}.
This error has been corrected in Fig.~\ref{figflucs}.}.

Finally, for the case of ideal hydrodynamics, 
analytic solutions to the hydrodynamic equations
are known \cite{Baym:1984sr,Chojnacki:2006tv,Nagy:2007xn}.
Specifically, the code for central 
collisions \cite{Baier:2006gy} has been found to agree with the results from
Ref.~\cite{Baym:1984sr} for ideal hydrodynamics. Since our code
agrees with Ref.~\cite{Baier:2006gy} in the case of central collisions
and when dropping the appropriate terms in the equations (\ref{maineq}),
this provides yet another test on our numerics.

To summarize, after conducting the above tests we are 
reasonably confident that our numerical
2+1 dimensional code solves the relativistic viscous hydrodynamic
equations (\ref{manyeq}) correctly.
This completes the setup of a viscous hydrodynamic description
of relativistic heavy-ion collisions. In the following sections,
we will review comparisons of viscous hydrodynamic simulations
to experimental data, for both Glauber and CGC initial conditions.

\section{Initial conditions: Glauber model vs. CGC}
\label{sec:two}

\subsection{The Glauber model}

In the Glauber model \cite{Kolb:2001qz}, the starting point is
the Woods-Saxon density distribution for nuclei,
\beq
\rho_A({\bf x})=\frac{\rho_0}{1+\exp{[(|{\bf x}|-R_0)/\chi]}},
\eeq
where for a gold nucleus with weight $A=197$ we use $R_0=6.4$ fm and
$\chi=0.54$ fm. The parameter $\rho_0$ is chosen such
that $\int d^3 {\bf x} \rho_A({\bf x})=A$.
One can then define the nuclear thickness function 
\beq
T_A(x^i)=\int_{-\infty}^\infty dz \rho_A({\bf x}) 
\eeq
and subsequently the number density of
nucleons participating in the collision ($n_{\rm Part}$)
and the number density of binary collisions ($n_{\rm Coll}$).
For a collision of two nuclei with weight A 
at an impact parameter $b$, one has
\bqa
n_{\rm Part}(x,y,b)&=&T_A\left(x+\frac{b}{2},y\right)
\left[1-\left(1-\frac{\sigma T_A\left(x-\frac{b}{2},y\right)}{A}\right)^A
\right]\nonumber\\
&&
+T_A\left(x-\frac{b}{2},y\right)
\left[1-\left(1-\frac{\sigma T_A\left(x+\frac{b}{2},y\right)}{A}\right)^A
\right],\nonumber\\
n_{\rm Coll}(x,y,b)&=&\sigma T_A\left(x+\frac{b}{2},y\right) T_A\left(x-\frac{b}{2},y\right),
\eqa
where $\sigma$ is the nucleon-nucleon cross section. We assume
\hbox{$\sigma\simeq 40$ mb} for Au+Au collisions at $\sqrt{s}=200$
GeV per nucleon pair.

While the total number of participating nucleons
$N_{\rm Part}(b)=\int dx dy n_{\rm Part}(x,y,b)$
will be used to characterize the centrality class of the 
collision, as an initial condition for the energy density
we will only use the parametrization
\beq
\epsilon(\tau=\tau_0,x,y,b)={\rm const}\times n_{\rm Coll}(x,y,b),
\label{edglauber}
\eeq
since it gives a sensible description of the multiplicity
distribution of experimental data, as will be discussed later on.
In the following, ``Glauber-model initial condition'' is
used synonymous to Eq.~(\ref{edglauber}).

The constant in Eq.~(\ref{edglauber}) is chosen such that the 
central energy density for zero impact parameter,
$\epsilon(\tau=\tau_0,0,0,0)$ corresponds to a predefined
temperature $T_i$ via the equation of state. 
This temperature will be treated as a free parameter and
is eventually fixed by matching to experimental data on
the multiplicity.

\subsection{The CGC model}

The other model commonly used to obtain initial conditions for 
hydrodynamics is the so-called Color-Glass-Condensate approach, 
based on ideas of gluon saturation at high energies.  In particular, 
we use a modified version of the KLN (Kharzeev-Levin-Nardi) 
$k_T$-factorization approach \cite{Kharzeev:2002ei}, 
due to Drescher {\it et al.} \cite{Drescher:2006pi}. We follow exactly 
the procedure described in \cite{Dumitru:2007qr} and in fact we use the same 
numerical code, provided to us by the authors and only slightly modified 
to output initial conditions suitable for input into our viscous 
hydrodynamics program.

In this model, the number density of gluons produced in a 
collision of two nuclei with atomic weight $A$ is given by
\beq
  \frac{dN_g}{d^2 {\bf x}_{T}dY} = {\cal N}
   \int \frac{d^2{\bf p}_T}{p^2_T}
  \int^{p_T} d^2 {\bf k}_T \;\alpha_s(k_T) \;
  \phi_A(x_1, ({\bf p}_T+{\bf k}_T)^2/4;{\bf x}_T)\;
              \phi_A(x_2, ({\bf p}_T - {\bf k}_T)^2/4;{\bf x}_T)
 \label{eq:ktfac}
\eeq
where ${\bf p}_T$ and $Y$ are the transverse momentum and 
rapidity of the produced gluons, respectively.  
$x_{1,2} = p_T\times\exp(\pm Y)/\sqrt{s}$ is the momentum fraction 
of the colliding gluon ladders with $\sqrt{s}$ the center of mass 
collision energy and $\alpha_s(k_T)$ is the strong coupling constant 
at momentum scale $k_T \equiv \left| {\bf k}_T \right|$.

The value of the normalization constant $\cal N$ is unimportant here, 
since as for Glauber initial conditions, we treat the overall normalization 
of the initial energy density distribution as a free parameter.
The unintegrated gluon distribution functions are taken as
\beq  \label{uGDF}
\phi (x,k^2_{T}; {\bf x}_{T}) =
\frac{1}{\alpha_s (Q^2_s)} \frac{Q^2_s}{\textrm{max}(Q^2_s,k^2_{T})}
\,P({\bf x}_{T})(1-x)^4~,
\eeq
$P({\bf x}_{T})$ is the probability of finding at least one nucleon at transverse position ${\bf x}_{T}$, taken from the definition for $n_{\rm Part}$
\beq
P({\bf x}_{T}) = 1-\left(1-\frac{\sigma T_A}{A}\right)^A,
\eeq
where $T_A$ and $\sigma$ are as defined in the previous section.

The saturation scale at a given
momentum fraction $x$ and transverse coordinate ${\bf x}_{T}$ is given by
\beq
  Q^2_{s}(x,{\bf x}_T) =
  2\,{\rm GeV}^2\left(\frac{T_A({\bf x}_T)/P({\bf x}_T)}{1.53/{\rm fm}^2}\right)
  \left(\frac{0.01}{x}\right)^\lambda~.
  \label{eq:qs}
\eeq
The growth speed is taken to be $\lambda = 0.288$.


The initial conditions for hydrodynamic evolution require that we specify 
the energy density in the transverse plane at some initial proper time 
$\tau_0$ at which the medium has thermalized.  Eq.~(\ref{eq:ktfac}), on the 
other hand, is in principle valid at a time $\tau_s = 1/Q_s$ at which the medium is likely not 
yet in thermal equilibrium.  To obtain the desired initial conditions, we 
again follow \cite{Dumitru:2007qr} and assume that the number of gluons is 
effectively conserved during the evolution from $\tau_s$ to $\tau_0$ and so 
the number density profile is the same at both times, scaled by the 
one-dimensional Bjorken expansion $n(\tau_0) = \frac {\tau_s}{\tau_0} n(\tau_s)$.  
The energy density can then be obtained from the number density through 
thermodynamic relations---it is proportional to the number density to the 4/3 power.  
Again, we take the overall normalization as a free parameter, so the initial 
energy density is finally given as
\beq
 \epsilon(\tau=\tau_0,{\bf x}_T,b)={\rm const}\times \left[ \frac{dN_g}{d^2 {\bf x}_{T}dY}({\bf x}_T,b)\right] ^{4/3}
\label{edCGC}
\eeq
where the number density is given by Eq.~(\ref{eq:ktfac}) evaluated at central rapidity $Y=0$.

As a final comment, it should be pointed out that the original version of the
CGC, the McLerran-Venugopalan model \cite{McLerran:1993ni,McLerran:1993ka}, 
differs from the KLN ansatz we used here, as will be discussed in the next-section.

\subsection{Spatial and momentum anisotropy}
\label{sec:aniso}

\begin{figure}[t]
\includegraphics[width=.45\linewidth]{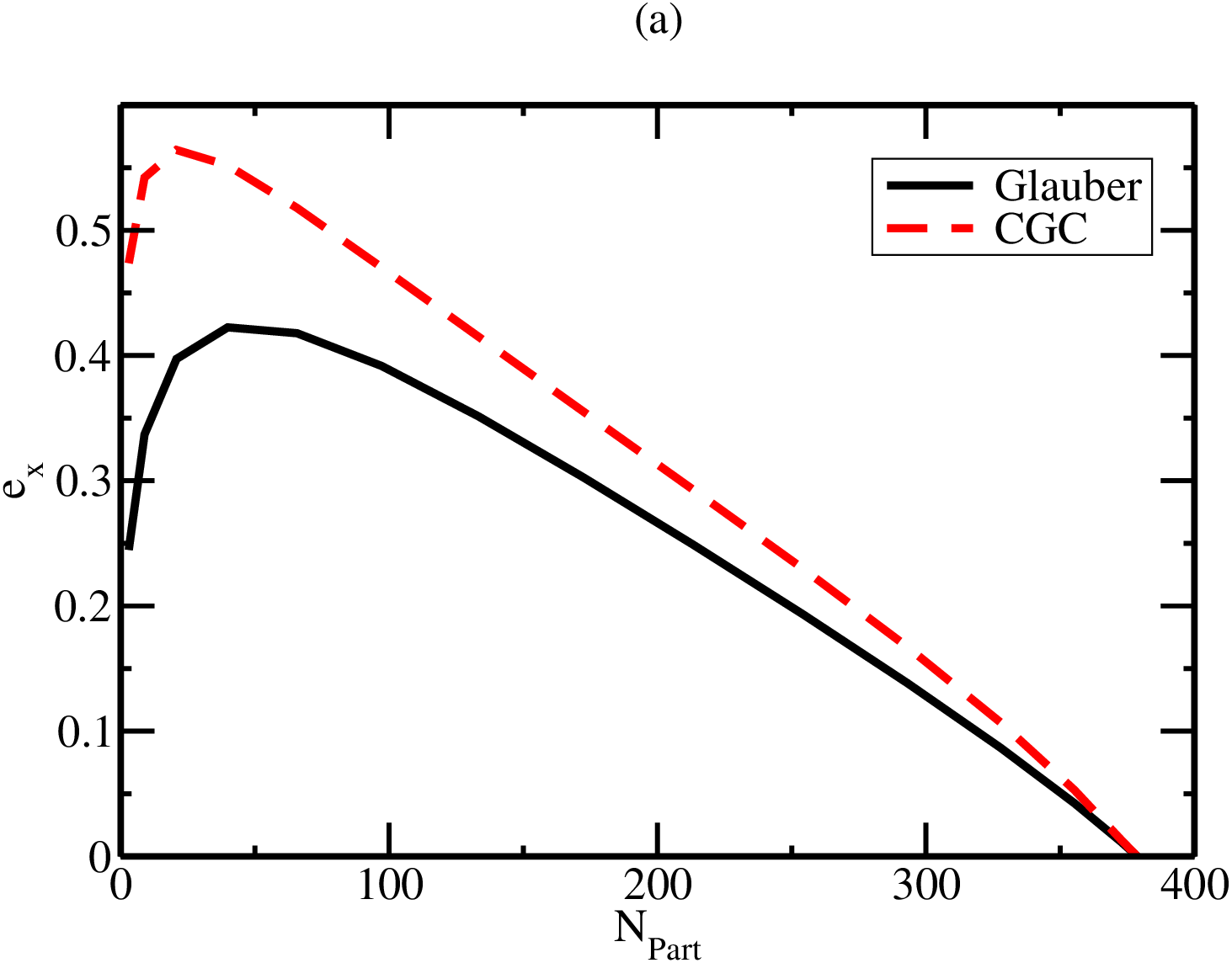}
\hfill
\includegraphics[width=.45\linewidth]{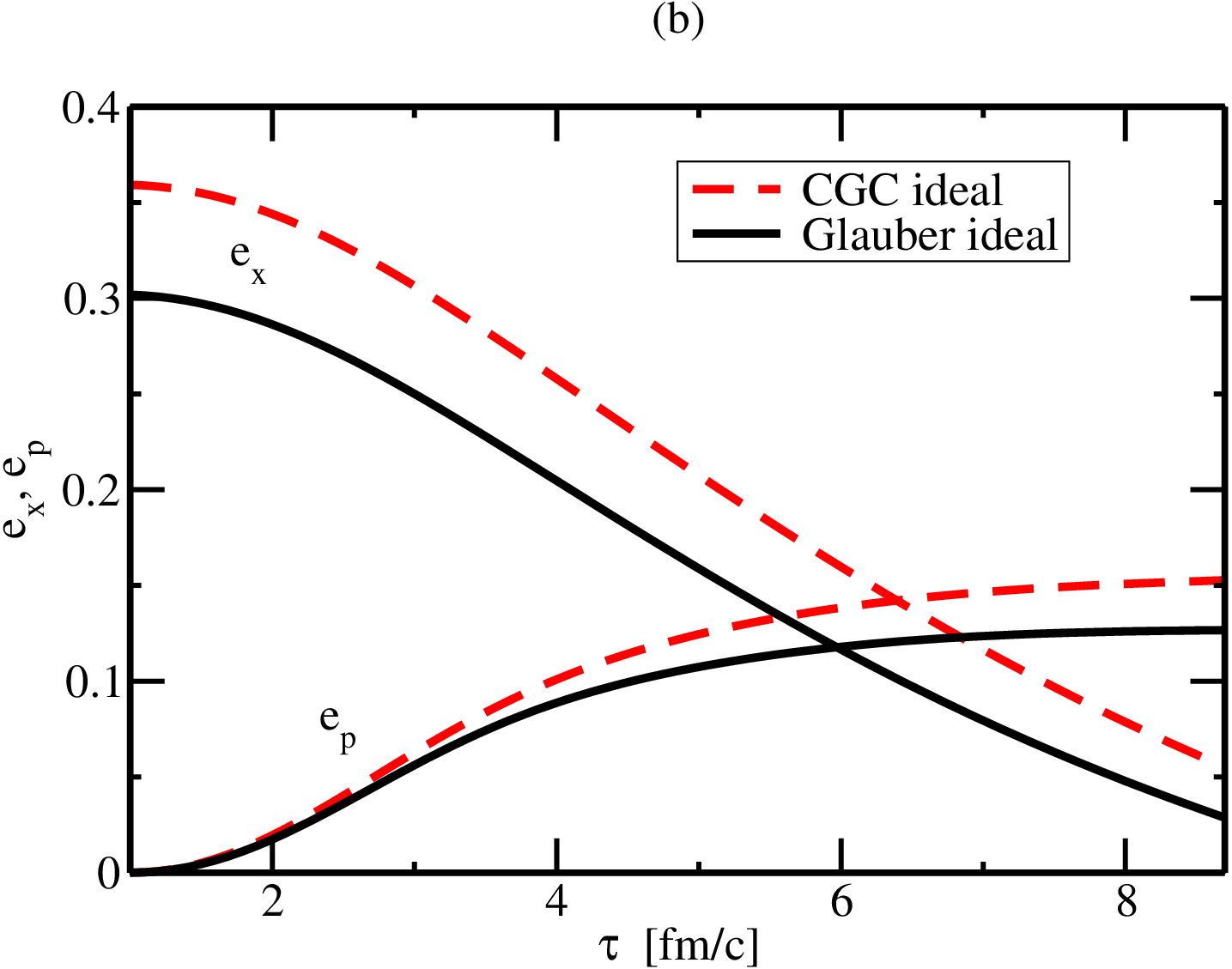}
\caption{(Color online) Left: The initial spatial anisotropy for the Glauber
and CGC model. Right: The time
evolution of the spatial and momentum anisotropy for a collision
with $b=7$ fm in ideal hydrodynamics.}
\label{fig:eccen}
\end{figure}

One of the key parameters discussed in the following is the
eccentricity (or spatial anisotropy) of the collision geometry.
Following \cite{Kolb:2001qz}, we define it as 
\beq
e_x\equiv 
\frac{\langle y^2-x^2\rangle_{\epsilon}}{\langle y^2+x^2\rangle_\epsilon},
\eeq
where $\langle \rangle_\epsilon$ denotes an averaging procedure over space 
with the energy density $\epsilon$ as a weighting factor. 
Shown in Fig.~\ref{fig:eccen}a, a plot of $e_x$ for different
centralities highlights the quantitative difference between the 
initial energy density from the Glauber and CGC model, 
Eq.~(\ref{edglauber}) and Eq.~(\ref{edCGC}), respectively.
As can be seen from this figure, the CGC model generally
gives a higher spatial anisotropy than the Glauber model.
Note that the results for the CGC model shown here are
extreme in the sense that the McLerran-Venugopalan model
gives spatial eccentricities which essentially match
the ones from the Glauber model \cite{Lappi:2006xc}.
This allows us to use the difference between the CGC and 
Glauber models as an indication of the systematic theoretical 
error stemming from the ignorance of the correct physical
initial condition.

Hydrodynamics converts pressure gradients into fluid velocities,
and hence one expects the spatial anisotropy to decrease at the expense
of a momentum anisotropy (which is related to the magnitude 
of the elliptic flow). We follow \cite{Kolb:1999it}
in defining a momentum anisotropy according to
\beq
e_p\equiv\frac{\langle T^{xx}-T^{yy}\rangle}{\langle T^{xx}+T^{yy}\rangle},
\label{momaniso}
\eeq
where we stress that here $\langle\rangle$ denotes spatial averaging
with weight factor unity. Fig.~\ref{fig:eccen}b shows the time
evolution in ideal hydrodynamics ($\eta/s\ll 1$)
of both the spatial and momentum anisotropies 
for a heavy-ion collision at $b=7$ fm modeled through 
Glauber and CGC initial conditions.
As one can see, for the same impact parameter,
the higher initial spatial anisotropy for the CGC
model eventually leads to a higher momentum anisotropy than
the Glauber model. Using a quasiparticle interpretation where
the energy momentum tensor is given by 
\beq
T^{\mu \nu}\propto \int \frac{d^3 {\bf p}}{(2 \pi)^3} \frac{p^\mu p^\nu}{E}
f\left(x^\mu, p^\mu\right),
\eeq
the momentum anisotropy $e_p$ can be approximately related to 
the integrated elliptic flow $v_2^{\rm int}(b)$,
with a proportionality factor of $\sim2$ \cite{Kolb:1999it,Ollitrault:1992bk}.
We find this proportionality to be maintained even for 
non-vanishing shear viscosity, as can be seen in Fig.~\ref{fig:v2int}.

\section{Results}
\label{sec:three}

\subsection{Which parameters matter?}

\begin{figure}[t]
\includegraphics[width=.45\linewidth]{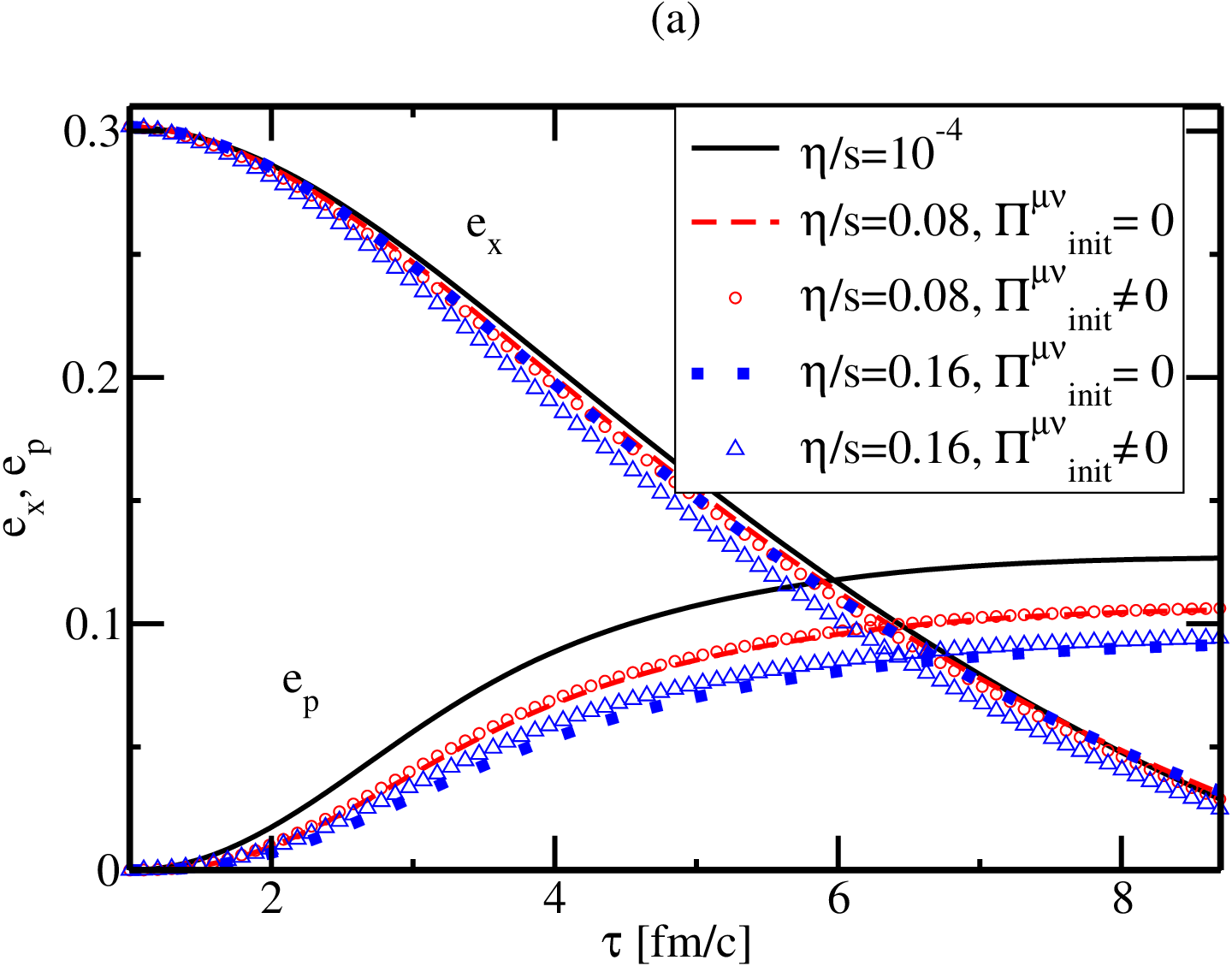}
\hfill
\includegraphics[width=.45\linewidth]{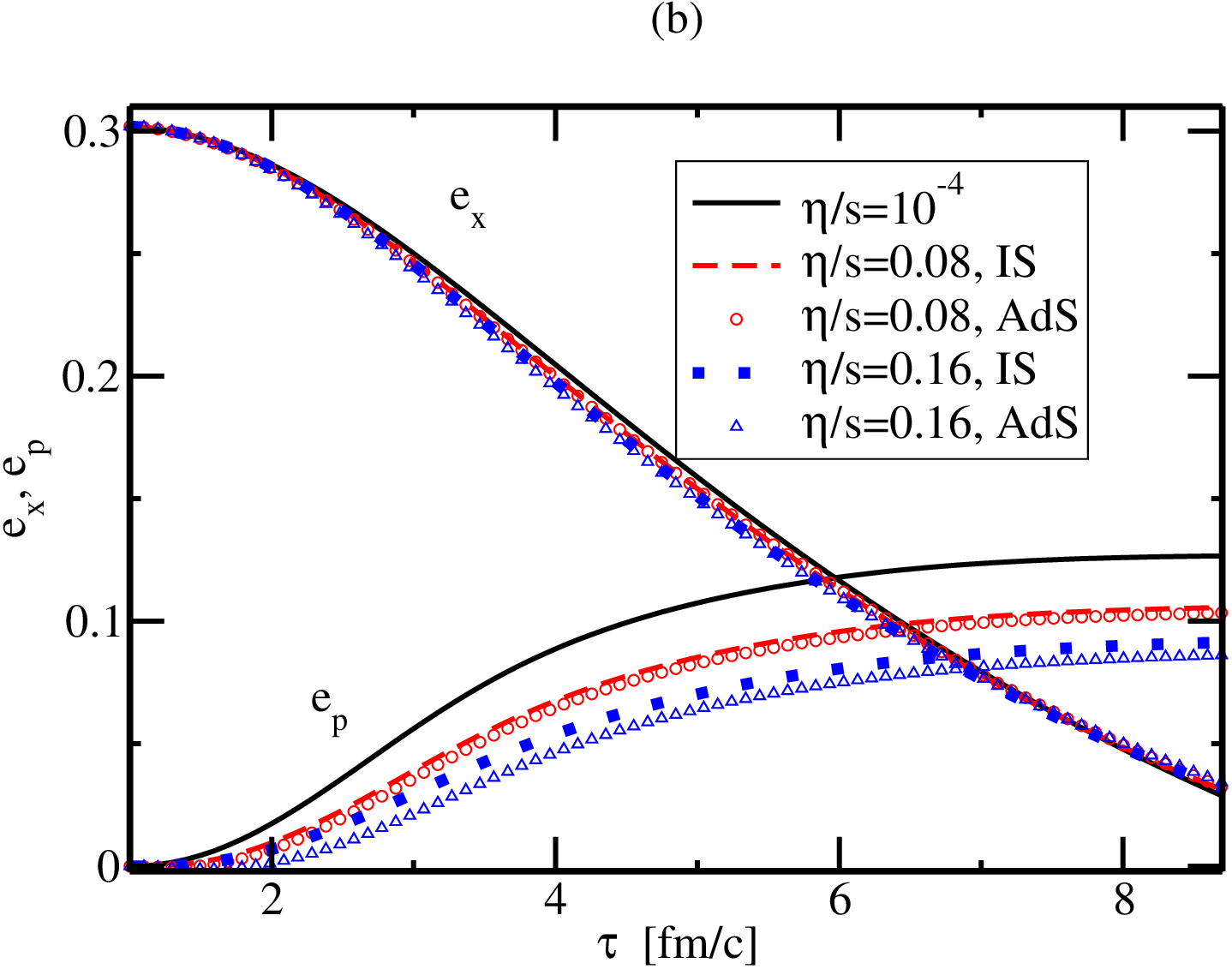}
\caption{(Color online) Spatial and momentum anisotropy for the Glauber model at $b=7$ fm
with $T_i=0.353$ GeV, $\tau_0=1$ fm/c and various values for the viscosity
(grid spacing $a=2\ {\rm GeV}^{-1}$). 
(a): The dependence on the 
initialization value of the shear tensor: shown are results
for vanishing initial value ($\Pi^{\mu\nu}_{\rm init}=0$) and 
Navier-Stokes initial value ($\Pi^{\mu\nu}_{\rm init}\neq 0$),
given in Eq.~(\protect\ref{FOvalue}).
(b): The dependence on the choice of value for $\tau_\Pi,\lambda_1$:
shown are results for $\tau_\Pi=\frac{6}{T}\frac{\eta}{s}$, 
$\lambda_1=0$ (labelled ``IS'') and 
$\tau_\Pi=\frac{2(2-\ln2)}{T}\frac{\eta}{s}$, 
$\lambda_1=\frac{\eta}{2 \pi T}$ (labelled ``AdS''). 
For $\tau_\Pi=\frac{2(2-\ln2)}{T}\frac{\eta}{s}$, 
the results for $\lambda_1=0$ (not shown) would be indistinguishable
by bare eye from those for $\lambda_1=\frac{\eta}{2 \pi T}$.}
\label{fignotmatter}
\end{figure}

In the following, we will attempt to obtain limits on
the mean value (throughout the hydrodynamic evolution) 
of the ratio $\eta/s$ from experimental data.
While e.g. temperature variations of $\eta/s$ are to
be expected in the real physical systems, probing for
such variations would invariable force us to introduce more
unknown parameters. We prefer to leave this program
for future studies once robust results for the mean value of $\eta/s$
exist.
Having fixed the equation of state and the freeze-out procedure
as explained in the previous sections, the remaining
choices that have to be made in the hydrodynamic model are the
\begin{itemize}
\item
Initial energy density profile: Glauber or CGC
\item
Initial value of shear tensor: vanishing or Navier-Stokes value
\item
Hydrodynamic starting time $\tau_0$
\item
Second-order coefficients: relaxation time $\tau_\Pi$ and $\lambda_1$
\item
Ansatz for non-equilibrium particle distribution Eq.~(\ref{fullfansatz})
\end{itemize}
where it is to be understood that we fix the initial energy density normalization ($T_i$) and the freeze-out temperature $T_f$ such that 
the model provides a reasonable description of the experimental data on 
multiplicity and $<p_T>$.
Historically, a strong emphasis has been put on requiring
a small value of $\tau_0$ for ideal hydrodynamics \cite{Kolb:2000sd,Heinz:2004pj}. 
For this reason, we will discuss the dependence on $\tau_0$ separately
in section \ref{notherm}.
A good indicator for which parameters
matter is the momentum anisotropy since it is very sensitive
to the value of $\eta/s$. From Fig.~\ref{fig:eccen} one therefore
immediately concludes that the choice of Glauber or CGC initial
conditions is important since it has a large effect on $e_p$.
Fortunately, most of the other choices turn out not to have a 
strong influence on the resulting $v_2$ coefficient and hence
the extracted $\eta/s$.  In the following we test for this sensitivity
by studying $e_p$ for a ``generic'' heavy-ion collision of two gold nuclei,
modeled by Glauber initial conditions at an initial starting
temperature of $T_i=0.353$, an impact parameter of $b=7$ fm,
and various choices of the above parameters. 

Fig.~\ref{fignotmatter} shows the time evolution of $e_x,e_p$
for various values of $\eta/s$. From these plots, it can be
seen that $e_p$ (and hence $v_2$) clearly is sensitive to the value of
$\eta/s$, suggesting that it can be a useful observable
to determine the viscosity of the fluid from experiment.
However, in order to be a useful probe of the fluid viscosity,
the dependence of the final value of $e_p$ on other parameters
should be much weaker than the dependence on $\eta/s$.
In Fig.~\ref{fignotmatter}a we show $e_p$, calculated
for $\Pi^{\mu \nu}(\tau_0)=0$ and $\Pi^{\mu \nu}(\tau_0)$ equal
to the Navier-Stokes value, Eq.~(\ref{FOvalue}).
As can be seen from this figure, the resulting
anisotropies are essentially independent of this choice, 
corroborating the finding in Ref.~\cite{Song:2007fn,Song:2007ux}.
Similarly, in Fig.~\ref{fignotmatter}b we show $e_p$
calculated in simulations where the values of the 
second-order transport coefficients were 
either those of a weakly-coupled M\"uller-Israel-Stewart theory ($\tau_\Pi=6 \frac{\eta}{sT}$,
$\lambda_1=0$) or those inspired by a strongly coupled
${\cal N}=4$ SYM plasma ($\tau_\Pi=2(2-\ln 2) \frac{\eta}{s T}$,
$\lambda_1=\frac{\eta}{2 \pi T}$). 
Again, the dependence
of $e_p$ on the choice of the values of $\tau_\Pi,\lambda_1$
can be seen to be very weak for the values of $\eta/s$ shown here. 
This result is in stark contrast to the findings of 
Ref.~\cite{Song:2007fn}, where a large sensitivity on the value
of $\tau_\Pi$ was found. However, recall that 
Ref.~\cite{Song:2007fn} used evolution equations that
differ from Eq.~(\ref{maineq}) and in particular
do not respect conformal invariance. As argued in
section \ref{section01}, it is therefore expected
to encounter anomalously large sensitivity to
the value of the second order transport coefficients.

\begin{figure}[t]
\begin{center}
\includegraphics[width=.5\linewidth]{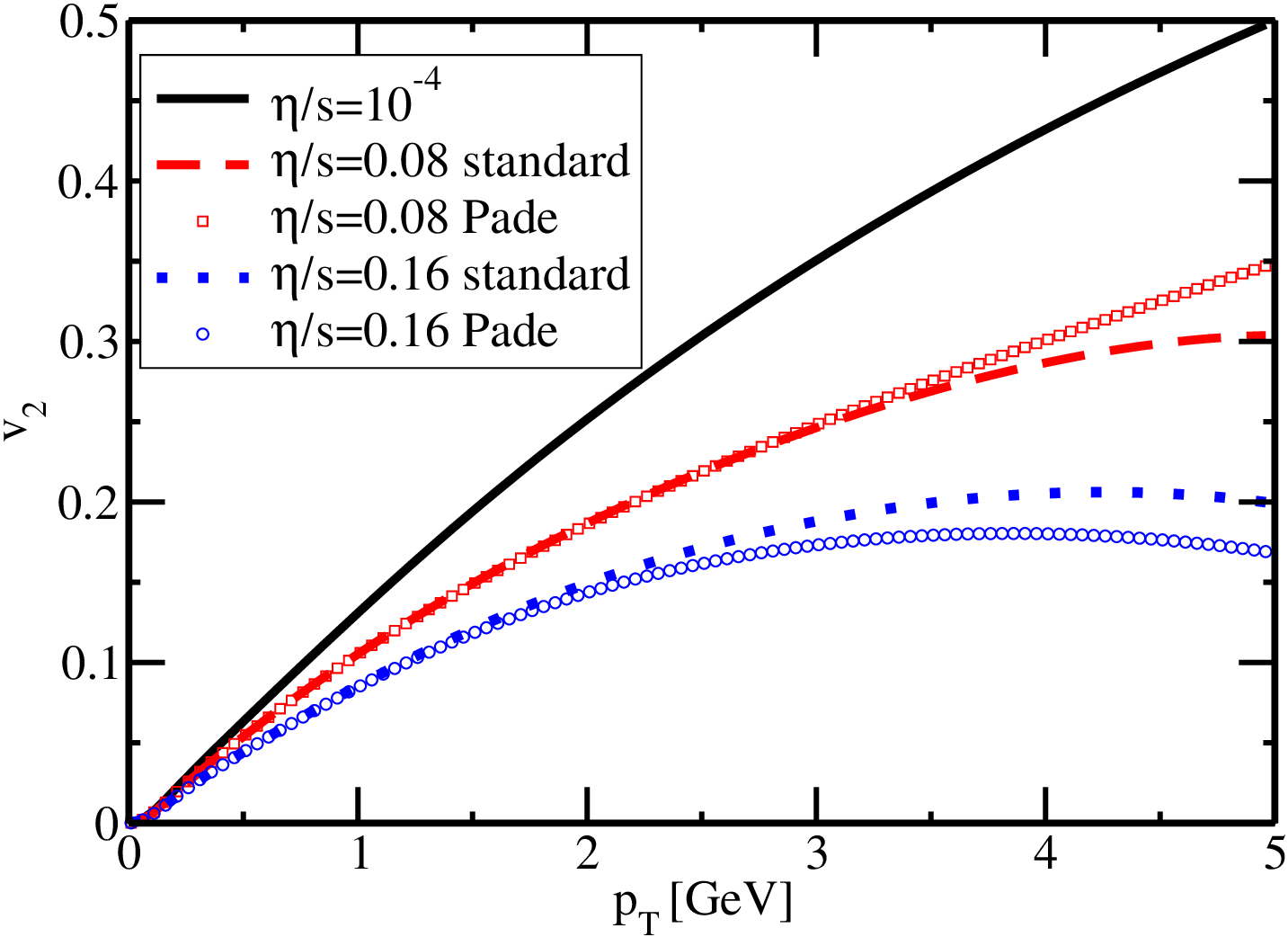}
\end{center}
\caption{(Color online) Charged hadron elliptic flow for the Glauber model at $b=7$ fm with $T_i=0.353$ GeV,
$\tau_0=1$ fm/c and various viscosities.}
\label{fig:Pade}
\end{figure}

To study the dependence of results on the ansatz of
the non-equilibrium particle distribution function (\ref{fullfansatz}),
one would want to quantify the effect of neglecting terms
of higher order in momenta in Eq.~(\ref{fullfansatz}). To estimate this,
let us rewrite $E d^3 N/d^3{\bf p}= E d^3 N^{(0)}/d^3{\bf p}+E d^3 N^{(1)}/d^3{\bf p}$,
where $N^{(0)}$ contains only the equilibrium part where 
$f(x^\mu,p^\mu)=f_0\left(\frac{p_\mu u^\mu}{T}\right)$, and perform a 
Pad\'e-type resummation,
\beq
E \frac{d^3 N^{\rm Pade}}{d^3{\bf p}}\equiv E \frac{d^3 N^{(0)}}{d^3{\bf p}}
\frac{1}{1-\frac{d^3 N^{(1)}}{d^3{\bf p}} \frac{d^3{\bf p}}{d^3 N^{(0)}}}.
\label{Padere}
\eeq
Since Eq.~(\ref{Padere}) contains powers of momenta to all orders when re-expanded,
the difference between the ansatz (\ref{fullfansatz}) and the Pad\'e resummed
particle spectra can give a handle on the systematic error of the truncation
used in Eq.~(\ref{fullfansatz}). Shown in Fig.~\ref{fig:Pade}, this
difference suggests that this systematic error is small 
for momenta $p_T\lesssim 2.5$ GeV. Therefore, we do not
expect our results to have a large systematic uncertainty coming from 
the particular ansatz (\ref{fullfansatz}) for these momenta.

To summarize, for values of $\eta/s\lesssim 0.2$, the 
results for the momentum anisotropy are essentially 
insensitive to the choices for the second-order transport
coefficients $\tau_\Pi,\lambda_1$ and the initialization
of the shear tensor $\Pi^{\mu\nu}(\tau=\tau_0)$.
Conversely, $e_p$ is sensitive to the value of viscosity
and the choice of initial energy density profile (initial eccentricity).
Since the physical initial condition is currently unknown, 
this dependence will turn out to be the dominant
systematic uncertainty in determining $\eta/s$ from 
experimental data.

\subsection{Multiplicity and radial flow}

As outlined in the introduction, we want to 
match the hydrodynamic model to experimental data
for the multiplicity, thereby fixing the constant in
Eqs.~(\ref{edglauber}),(\ref{edCGC}). This translates to fixing an initial
central temperature $T_i$ for $b=0$, which we will quote 
in the following. 

For a constant speed of sound, the evolution for ideal
hydrodynamics is isentropic, while for viscous
hydrodynamics additional entropy is produced.
Since the multiplicity is a measure of the entropy of
the system, one expects an increase of multiplicity
for viscous compared to ideal hydrodynamic evolution.
This increase in final multiplicity has been measured as a 
function of $\eta/s$ for the semi-realistic speed of 
sound Fig.~\ref{figcs2} in central heavy-ion collisions 
in Ref.~\cite{Romatschke:2007jx},
and found to be approximately\footnote{
The quoted fraction is for a hydrodynamic starting time of $\tau_0=1$ fm/c.
Reducing $\tau_0$ leads to considerably larger entropy production.} 
a factor of $0.75 \eta/s$.
(See Ref.~\cite{Lublinsky:2007mm,Dumitru:2007qr} for 
related calculations in simplified models.)
Reducing $T_i$ accordingly therefore ensures that
for viscous hydrodynamics, 
the multiplicity in central collisions will stay close
to that of ideal hydrodynamics.

Hydrodynamics gradually 
converts pressure gradients into flow velocities,
which in turn relate to the mean particle momenta.
Starting at a predefined time $\tau_0$ and requiring the hydrodynamic model spectra to
match the experimental data on particle $<p_T>$ 
then fixes the freeze-out temperature $T_f$.

\begin{figure}[t]
\includegraphics[width=.48\linewidth]{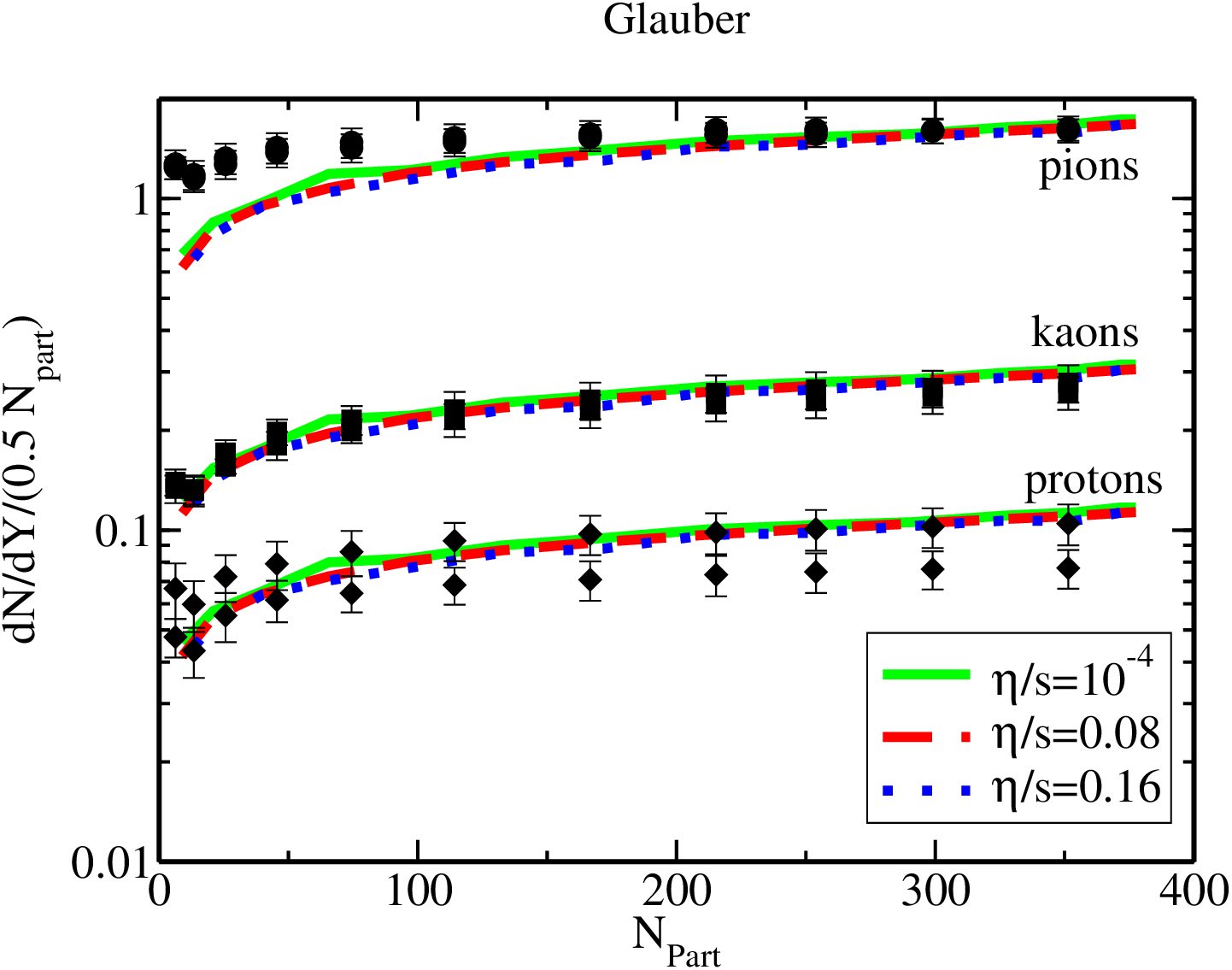}
\hfill
\includegraphics[width=.48\linewidth]{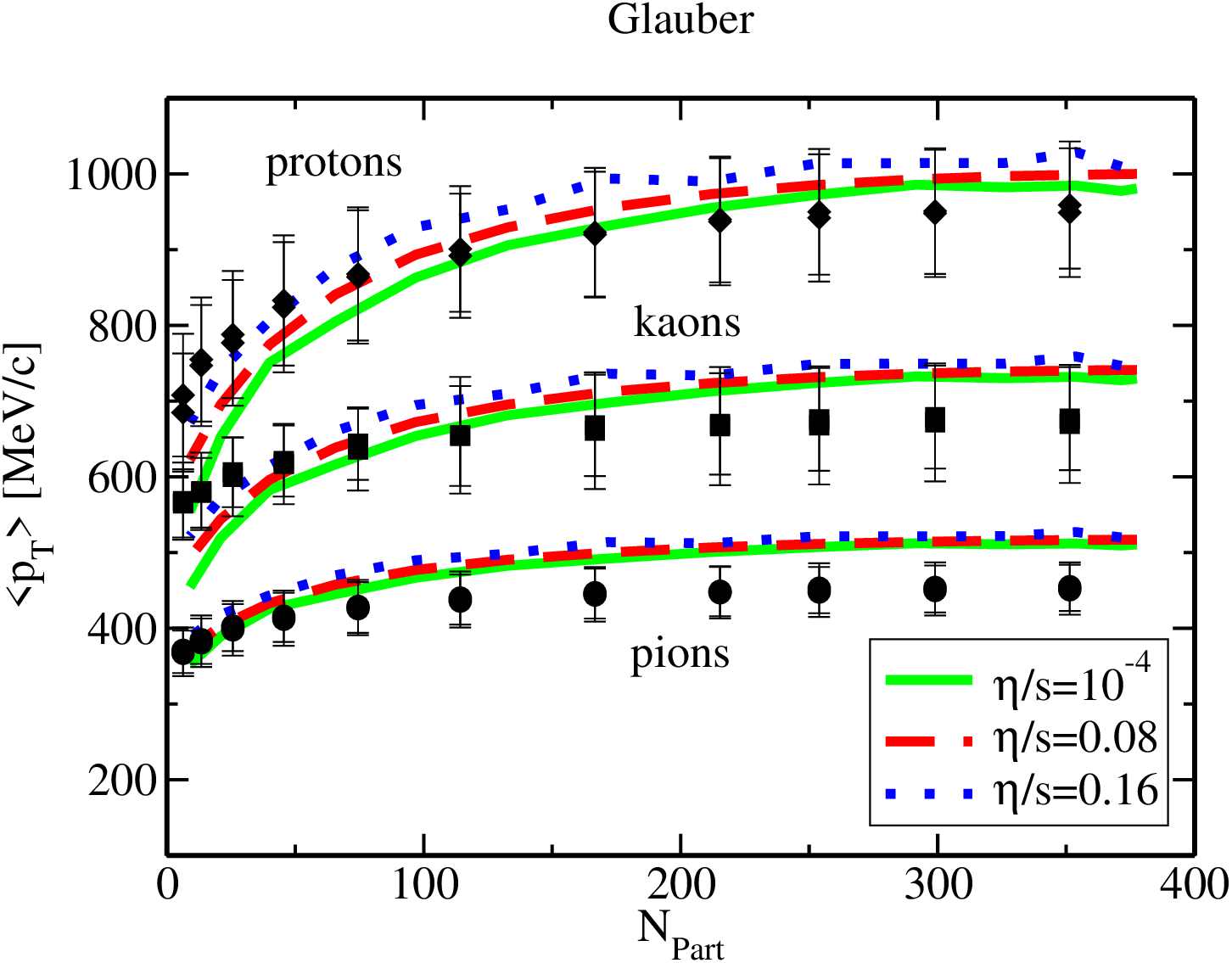}
\hfill\\
\vspace*{1cm}
\includegraphics[width=.48\linewidth]{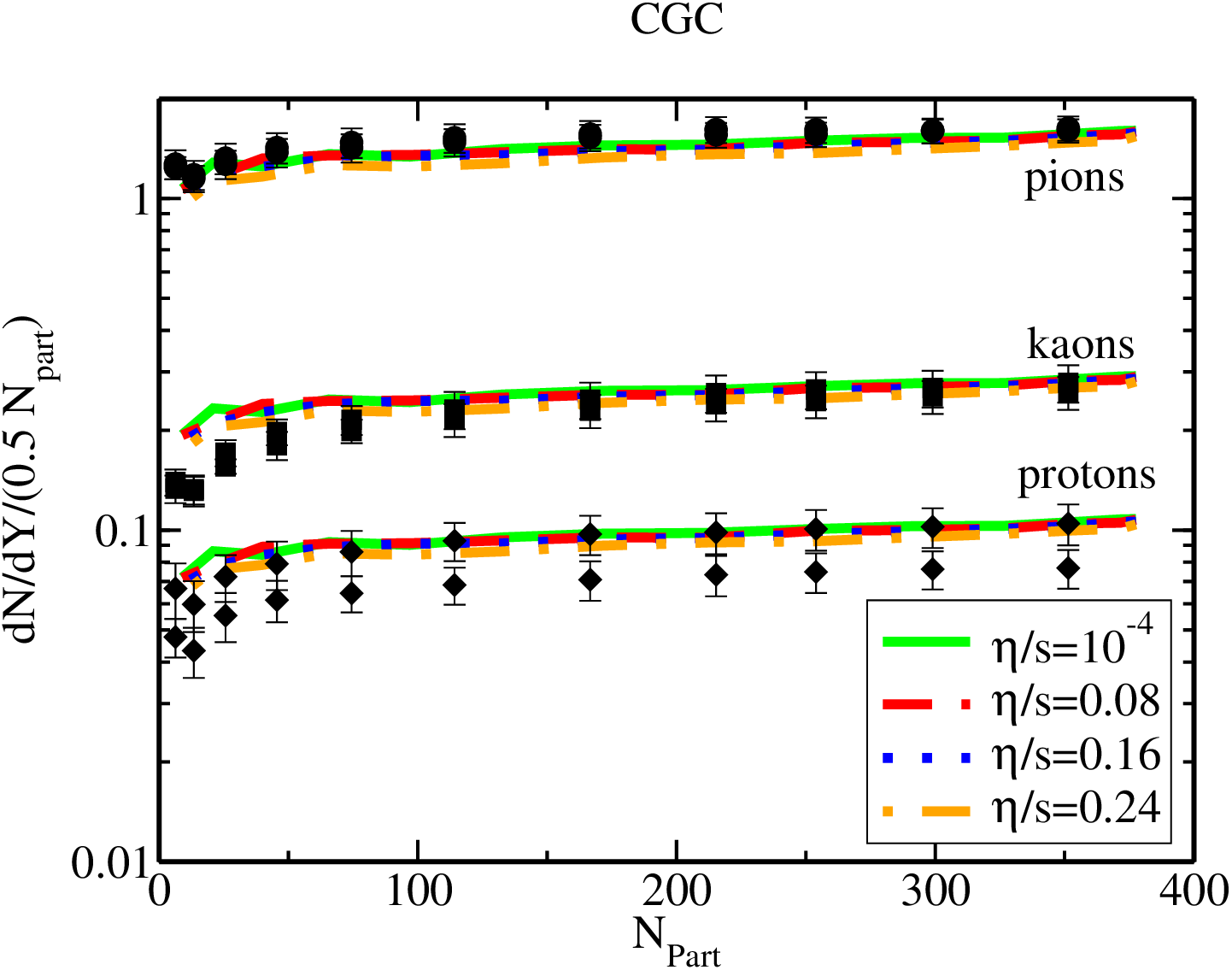}
\hfill
\includegraphics[width=.48\linewidth]{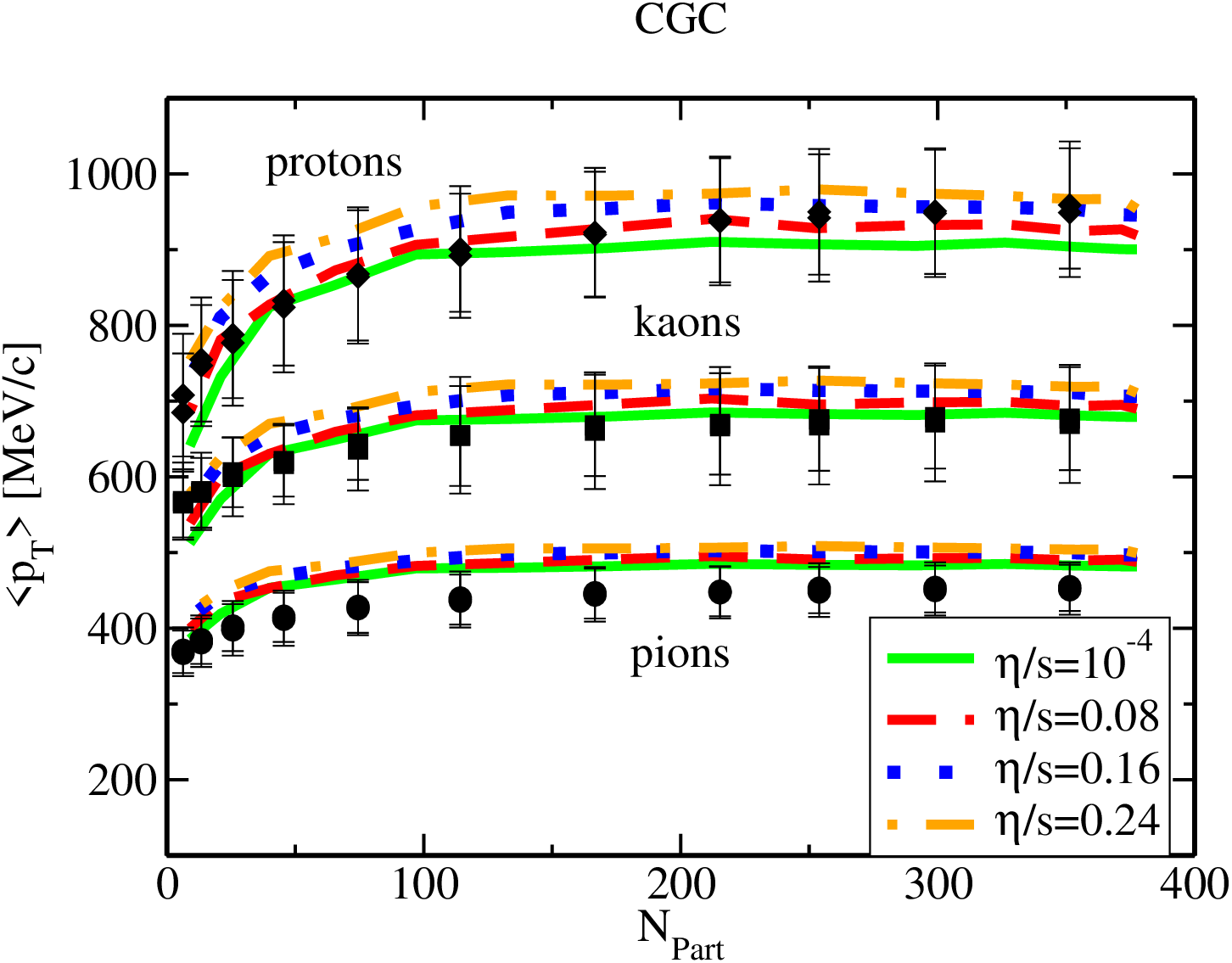}

\caption{(Color online) Centrality dependence of total
multiplicity $dN/dY$ and $<p_T>$ 
for $\pi^+,\pi^-,K^+,K^-,p$ and $\bar{p}$ 
from PHENIX \cite{Adler:2003cb} 
for Au+Au collisions at $\sqrt{s}=200$ GeV, compared to 
the viscous hydrodynamic model and various $\eta/s$, 
for Glauber initial conditions 
and CGC initial conditions. The model parameters used here are 
$\tau_0=1$ fm/c, $\tau_\Pi=6 \eta/s$,
$\lambda_1=0$, $T_f=140$ MeV 
and adjusted $T_i$ (see Table \ref{tab:par}). }
\label{mult1}
\end{figure}

For both Glauber-type and CGC-type model initial conditions, 
the experimental impact parameter 
dependence of the multiplicity and $<p_T>$
is reasonably well parametrized for both ideal hydrodynamics
as well as viscous hydrodynamics provided $T_i$ is adjusted
accordingly (see Fig.~\ref{mult1}). 
The values for $T_i$ used in the simulations
are compiled in Table~\ref{tab:par}.
We recall that no chemical potential is included in our
equation of state, prohibiting a distinction between particles
and anti-particles, and chemical and kinetic freeze-out of
particles occurs at the same temperature.
Furthermore, approximating the equilibrium particle-distributions
for bosons by a Boltzmann distribution (\ref{fullfansatz}) 
leads to small, but consistent underestimation of the multiplicity
of light particles, such as pions.
For these reasons, it does not make sense to attempt a precision fit 
to experimental data, especially for pions and protons. 
Rather, we have aimed for a sensible description of the 
overall centrality dependence of multiplicity and $<p_T>$ of kaons.

Note that in particular for the CGC model one could achieve a better fit to
the data on mean $<p_T>$ by increasing the freeze-out temperature
by $\sim 10$ MeV. This would also lead to a decrease in elliptic flow
for this model. However, to facilitate comparison between the CGC and 
Glauber initial conditions, we have kept $T_f$ the same for both models.

\begin{table}
\begin{center}
\begin{tabular}{|c|c|c|c|c|c|}
\hline
Initial condition & 
$\eta/s$ & 
$T_i$ [GeV] & 
$T_f$ [GeV] &
$\tau_0$ [fm/c] &  
a [GeV$^{-1}$]\\
\hline
Glauber& $10^{-4}$ & 0.340 & 0.14&1&2\\
Glauber& $0.08$ & 0.333 & 0.14&1&2\\
Glauber& $0.16$ & 0.327 & 0.14&1&2\\
CGC& $10^{-4}$ & 0.310 & 0.14&1&2\\
CGC& $0.08$ & 0.304 & 0.14&1&2\\
CGC& $0.16$ & 0.299 & 0.14&1&2\\
CGC& $0.24$ & 0.293 & 0.14&1&2\\
%
%
\hline
\end{tabular}
\end{center}
\caption{Summary of parameters used for the viscous hydrodynamics
simulations
}
\label{tab:par}
\end{table}

\subsection{Elliptic flow}

\begin{figure}[t]
\includegraphics[width=.445\linewidth]{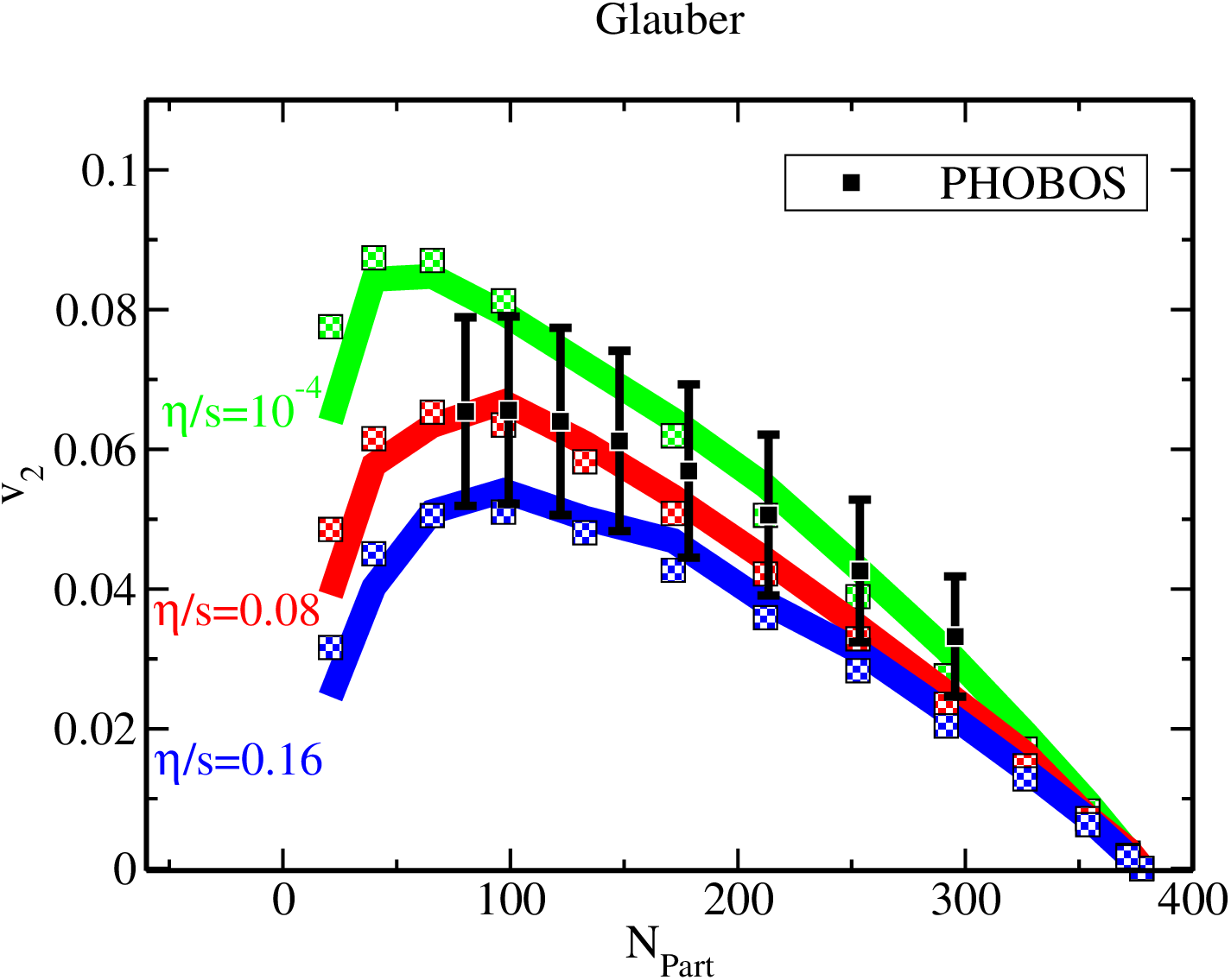}
\hfill
\includegraphics[width=.49\linewidth]{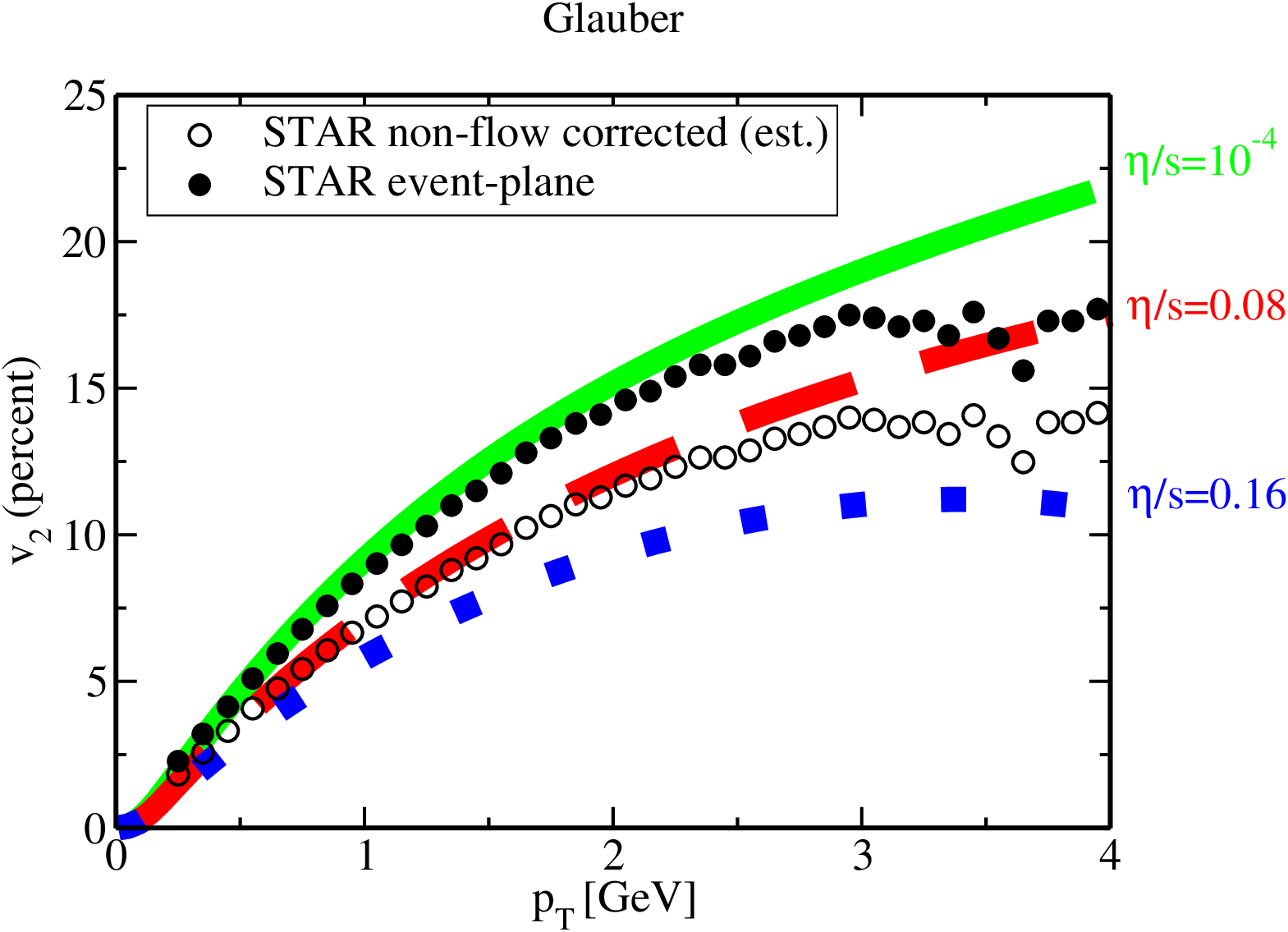}
\hfill\\
\vspace*{1cm}
\includegraphics[width=.445\linewidth]{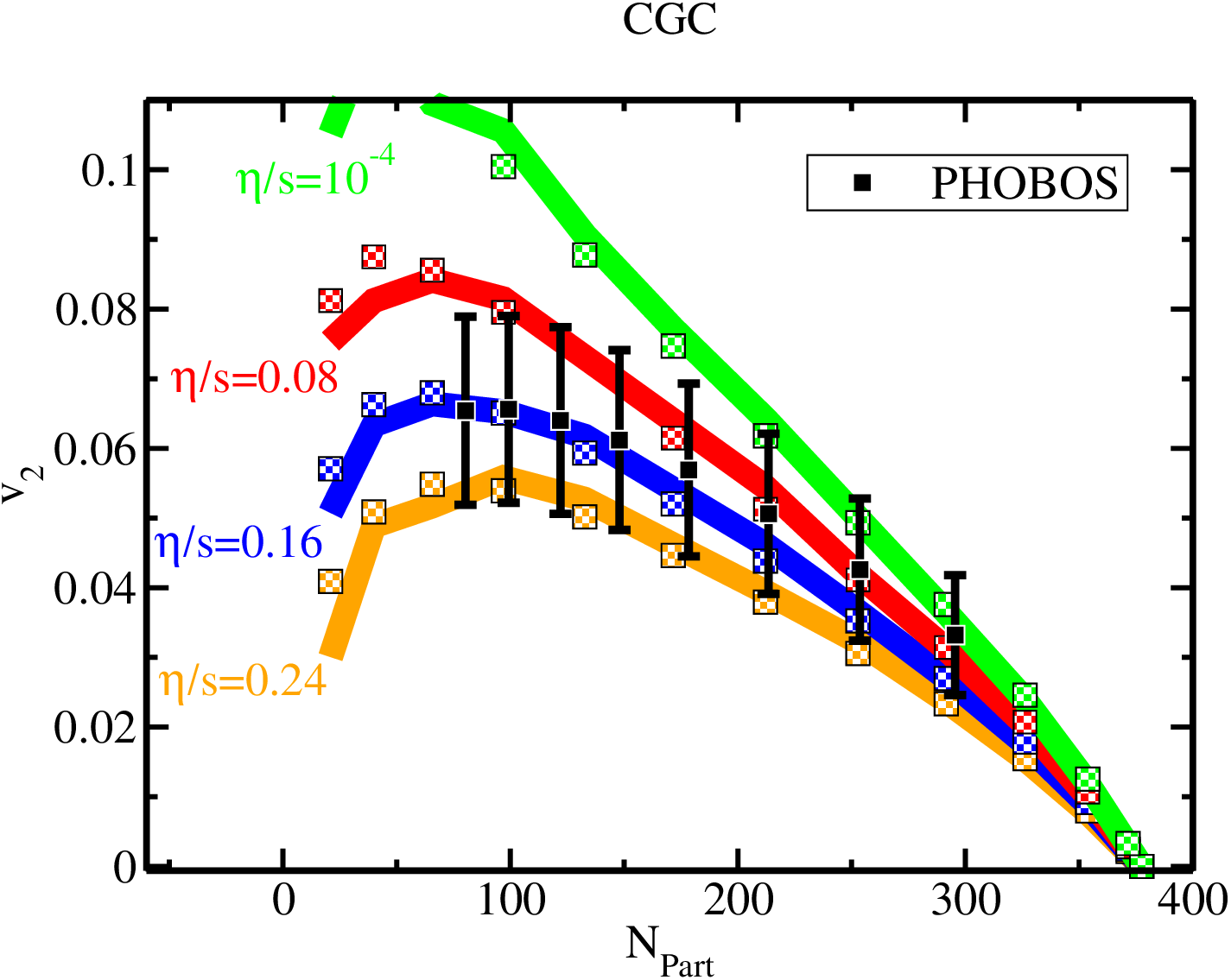}
\hfill
\includegraphics[width=.49\linewidth]{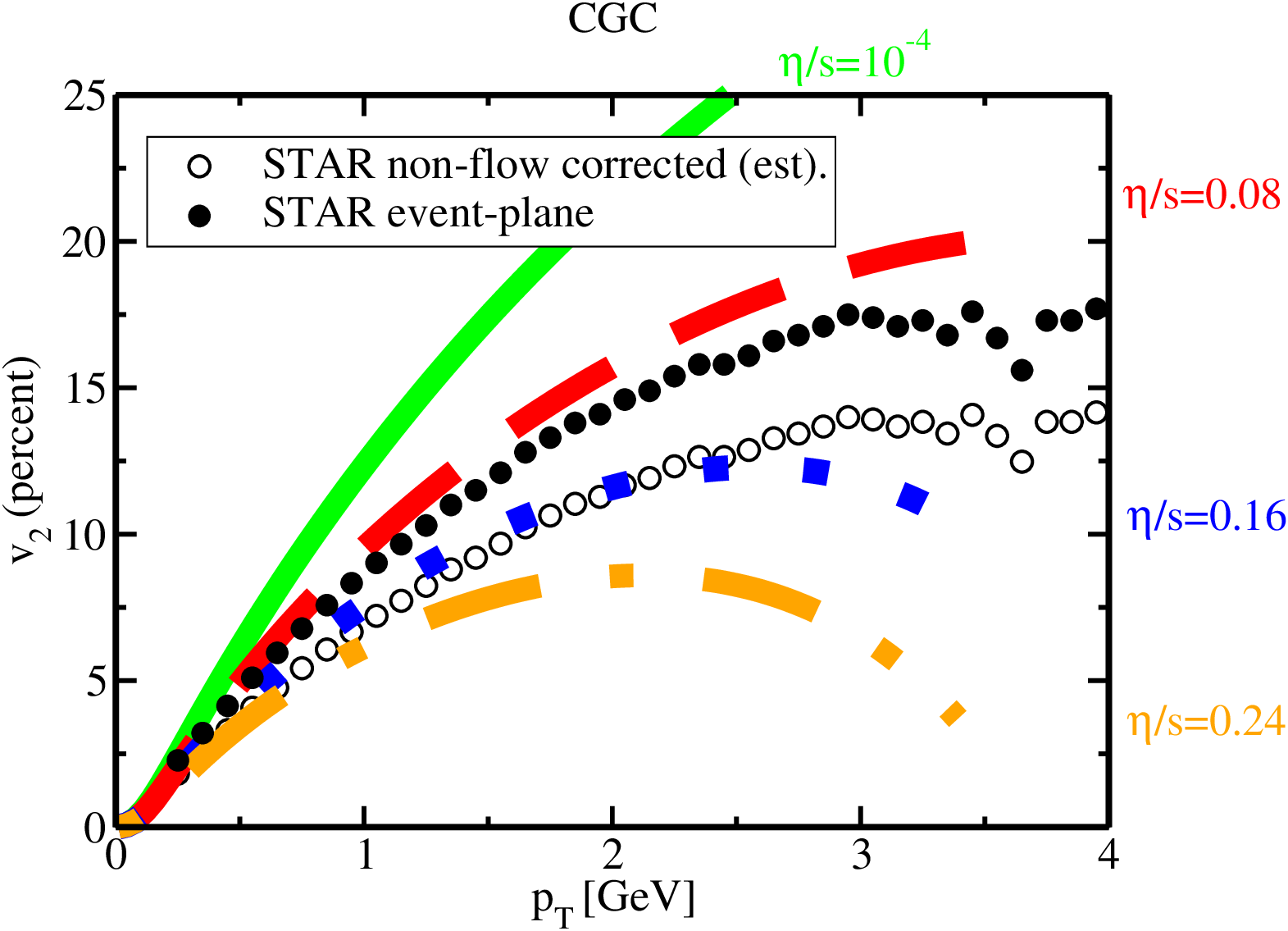}
\hfill

\caption{(Color online) Comparison of hydrodynamic models to
experimental data on charged hadron integrated (left)
and minimum bias (right) elliptic flow by PHOBOS \cite{Alver:2007qw} 
and STAR \cite{Abelev:2008ed}, respectively.
STAR event plane data has been reduced by 20 percent 
to estimate the removal of non-flow contributions \cite{Abelev:2008ed,Poskanzer}.
The line thickness for the hydrodynamic model curves is 
an estimate of the accumulated numerical error (due to, e.g., finite
grid spacing). The integrated $v_2$ coefficient 
from the hydrodynamic models
(full lines) is well reproduced by $\frac{1}{2} e_p$ (dots); indeed,
the difference between the full lines and dots gives an
estimate of the systematic uncertainty of the freeze-out prescription.}
\label{fig:v2int}
\end{figure}

Having fixed the parameters $\tau_0,T_i,T_f$ for a given $\eta/s$
to provide a reasonable description of the experimental data,
a sensible comparison between the model and experimental results
for the elliptic flow coefficient can be attempted.
For charged hadrons, the integrated and minimum-bias $v_2$ coefficients 
are shown in Fig.~\ref{fig:v2int} 
for Glauber and CGC initial conditions.
As noted in section \ref{sec:aniso}, charged hadron
$v_2^{\rm int}$ turns out to be very well reproduced by
the momentum eccentricity $\frac{1}{2}\ e_p$, evaluated when
the last fluid cell has cooled below $T_f$. This agreement
is independent from impact parameter or viscosity
and hence may serve as a more direct method on obtaining 
an estimate for $v_2^{\rm int}$
if one cannot (or does not want to) make use of the 
Cooper-Frye freeze-out procedure described in
section \ref{sec:fo}.

The comparison of the hydrodynamic model 
to experimental data with $90$\% confidence level
systematic error bars from PHOBOS \cite{Alver:2007qw} 
for the integrated elliptic flow in Fig.~\ref{fig:v2int}
suggests a maximum value of $\eta/s\sim0.16$ for Glauber-type
and $\eta/s\sim0.24$
for CGC-type initial conditions.
Whereas for Glauber initial conditions, ideal hydrodynamics 
($\eta/s\sim0$) gives results consistent with PHOBOS data,
for CGC initial conditions zero viscosity does not give a good
fit to the data, which is consistent with previous findings 
\cite{Hirano:2005xf}.

For minimum-bias $v_2$, to date only experimental data 
using the event-plane method are available, 
where the statistical, but not the systematic error 
of that measurement is directly accessible.
The dominant source of systematic error is associated
with the presence of so-called non-flow effects \cite{Ollitrault:1995dy}.
Recent results from STAR suggest that removal of 
these non-flow effects imply a reduction of the event-plane minimum
bias $v_2$ by $20$ percent \cite{Abelev:2008ed,Poskanzer}.
For charged hadrons, a comparison of both the event-plane and the 
estimated non-flow corrected experimental data from STAR
with the hydrodynamic model is shown in Fig.~\ref{fig:v2int}.

For Glauber-type initial conditions, the data on
minimum-bias $v_2$ for charged hadrons is consistent with 
the hydrodynamic model for viscosities in the range
$\eta/s\in[0,0.1]$, while for the CGC case the respective
range is $\eta/s\in[0.08,0.2]$. It is interesting
to note that for Glauber-type initial conditions,
experimental data for both the integrated as well as
the minimum-bias elliptic flow coefficient (corrected for non-flow effects)
seem to be reproduced best\footnote{
In Ref.~\cite{Romatschke:2007mq} a lower value of $\eta/s$
for the Glauber model was reported. 
The results for viscous hydrodynamics shown in Fig.~\ref{fig:v2int} 
are identical
to Ref.~\cite{Romatschke:2007mq}, but the new STAR data with non-flow corrections
became available only after \cite{Romatschke:2007mq} had been
published.} by a hydrodynamic model with
$\eta/s=0.08\simeq\frac{1}{4\pi}$. This number has
first appeared in the gauge/string duality context 
\cite{Policastro:2001yc} and has
been conjectured to be the universal lower bound
on $\eta/s$ for any quantum field theory at finite temperature
and zero chemical potential \cite{Kovtun:2004de}. 
For CGC-type initial conditions, 
the charged hadron
$v_2$ data seems to favor a hydrodynamic model with
$\eta/s\sim 0.16$, well above this bound.

\subsection{Early vs. late thermalization}
\label{notherm}

\begin{figure}[t]
\includegraphics[width=.47\linewidth]{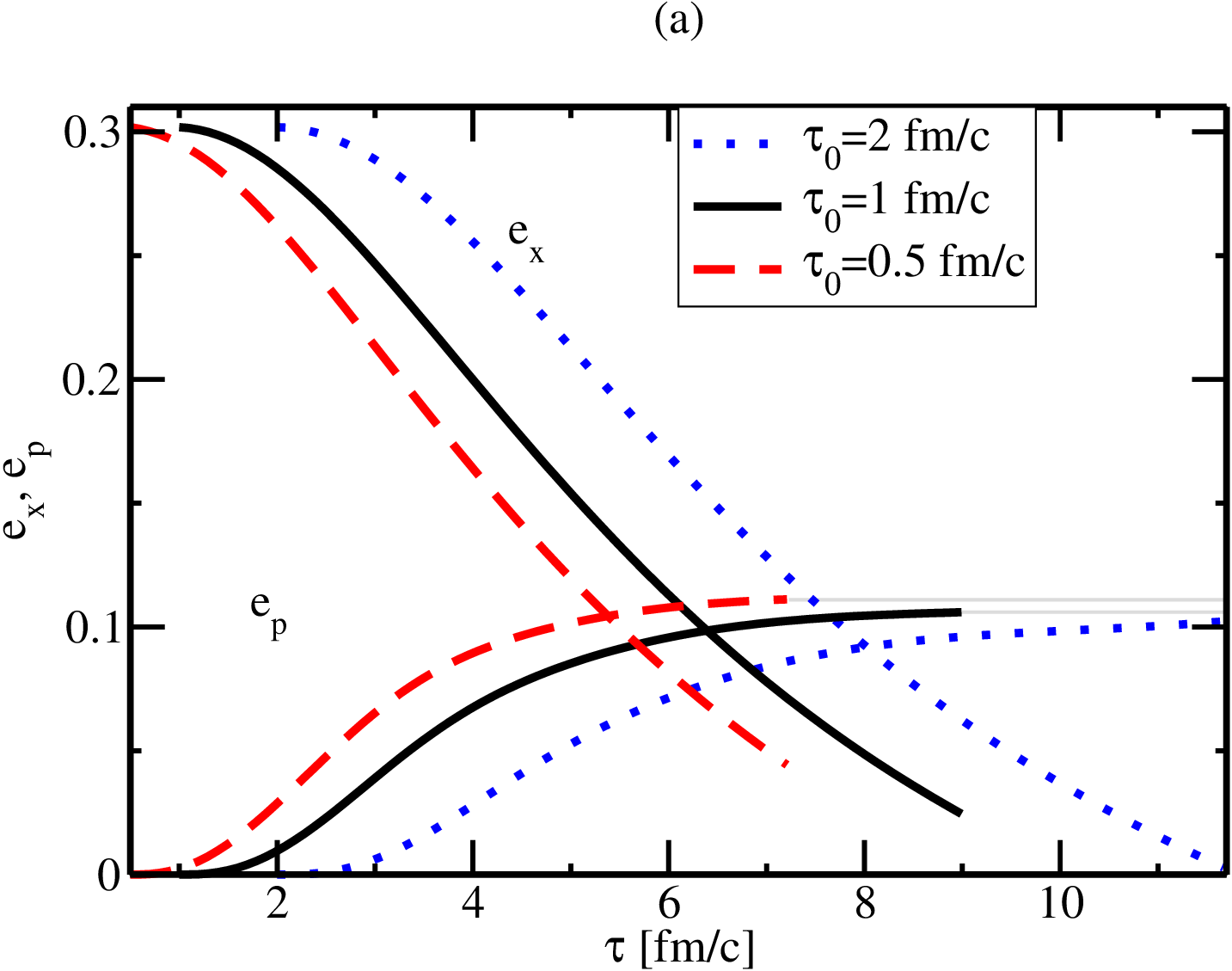}
\hfill
\includegraphics[width=.44\linewidth]{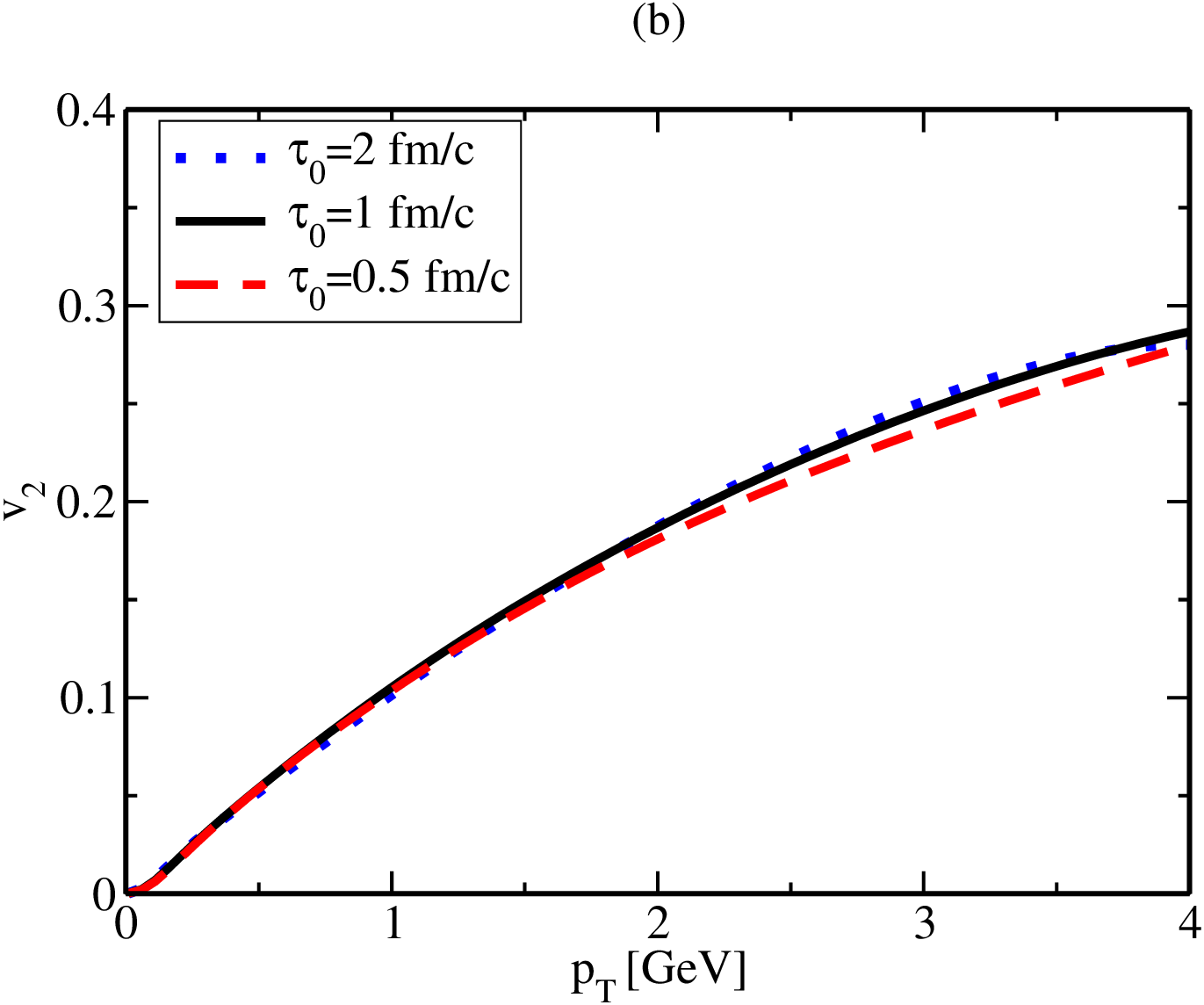}
\caption{(Color online) Momentum anisotropy (a) and elliptic flow for charged hadrons (b)
for $b=7$ fm, $\eta/s=0.08$ and different hydrodynamic initialization times $\tau_0$.
Horizontal light gray lines in (a) are visual aids to compare the final value of 
$e_p$. As can be seen from these plots, neither the final $e_p$ nor 
the charged hadron $v_2$ depend sensitively on the value of $\tau_0$
if the same energy distribution is used as initial condition at the respective
initialization times. Simulation parameters were 
$T_i=0.29\ {\rm GeV},\, T_f=0.14\ {\rm GeV}$ for $\tau_0=2$ fm/c,
$T_i=0.36\ {\rm GeV},\, T_f=0.15\ {\rm GeV}$ for $\tau_0=1$ fm/c,
and $T_i=0.43\ {\rm GeV},\, T_f=0.16\ {\rm GeV}$ for $\tau_0=0.5$ fm/c.
}
\label{fig:tdep}
\end{figure}

Currently, there seems to be a common misunderstanding in the
heavy-ion community that hydrodynamic models 
can universally only reproduce experimental data if they are initialized at 
early times $\tau_0<1$ fm/c. This notion has been labeled 
``early thermalization'' and continues to create a lot of confusion.
In this section, we argue that the matching 
of hydrodynamics to data itself does not require $\tau_0<1$ fm/c.
It is the additional assumptions about pre-equilibrium dynamics 
that lead to this conclusion for the Glauber initial conditions.

Performing hydrodynamic simulations in the way we have described earlier,
the energy density distribution is specified by either the Glauber or CGC
model at an initial time $\tau_0$. In Fig.~\ref{fig:tdep} we show the result for the 
elliptic flow coefficient (or the momentum anisotropy) for three different
values of $\tau_0$, namely $0.5,1$ and $2$ fm/c, where also $T_i$ and $T_f$
have been changed in order to obtain roughly the same multiplicity and
mean $p_T$ for each $\tau_0$. As can be seen from this figure,
the resulting final elliptic flow coefficient is essentially independent
of the choice of $\tau_0$. In particular, this implies that experimental
data for bulk quantities 
can be reproduced by hydrodynamic models also for large initialization times,
so no early thermalization assumption is needed.

\begin{figure}[t]
\begin{center}
\includegraphics[width=.5\linewidth]{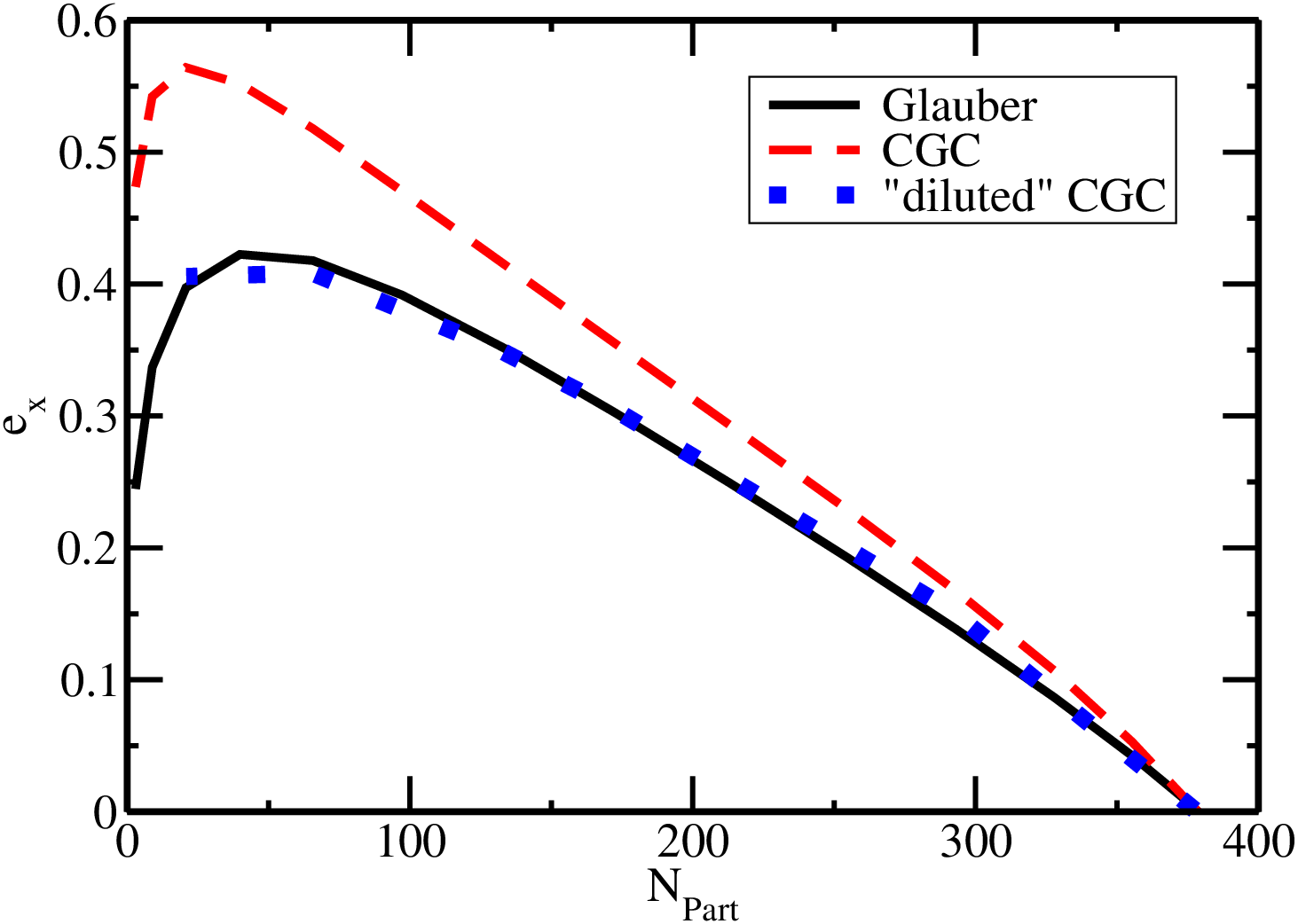}
\end{center}
\caption{(Color online) Spatial eccentricity for the Glauber and CGC model compared
to evolving the CGC model according to Eq.(\protect\ref{dilution}) 
for $\tau=1.5$ fm/c. This implies that starting with Glauber-type initial
conditions at $\tau_0>1$ fm/c may not be unreasonable.}
\label{fig:dilu}
\end{figure}

However, it is true that the above procedure assumes that the energy
density distribution remains unchanged up to the starting time
of hydrodynamics, which arguably becomes increasingly inaccurate for larger $\tau_0$.
It has therefore been suggested \cite{Kolb:2000sd}
to mimic the pre-hydro time evolution of the
energy density distribution by assuming free-streaming of partons.
Assuming free-streaming gives the maximal contrast to assuming hydrodynamic evolution,
since the latter corresponds to very strong interactions while the former corresponds
to no parton interactions at all. Indeed, one can calculate the 
effect of the free-streaming evolution on the spatial anisotropy, 
finding \cite{Kolb:2000sd}
\beq
e_x(\tau)=\frac{e_x(0)}{1+\frac{\tau^2}{3 <R^2>}},\quad
<R^2>=\frac{\int d^2 {\bf x} \epsilon(\tau=0)}{\int d^2 {\bf x} \frac{(x^2+y^2)}{2}\epsilon(\tau=0)}.
\label{dilution}
\eeq
This implies that the spatial anisotropy decreases with time, whereas one
can show that free-streaming does not lead to a build-up of $e_p$. In other words,
the eccentricity gets diluted without producing elliptic flow, such that 
once hydrodynamic evolution starts, it will not lead to as much $v_2$
as it would have without the dilution effect\footnote{
It seems that if one forces the energy-momentum tensor at the end of
free-streaming period to match to that of \emph{ideal} hydrodynamics
(instantaneous thermalization), the resulting fluid velocities
are anisotropic, i.e. correspond to a non-vanishing elliptic flow coefficient
\cite{Pasi,Broniowski:2008vp}. It is possible that this effect stems
from neglecting velocity gradients (viscous hydrodynamic corrections) 
in the matching process. We ignore the complications of the detailed
matching from free-streaming to hydrodynamics in the following.}. 
It is tempting to
conclude from this that by comparing to experimental data on elliptic flow 
one could place an upper bound on the maximally allowed dilution time,
and interpret this as the thermalization time of the system.
One should be aware, however, that this bound will depend
on the assumption made about the pre-hydro evolution. Furthermore, 
one should take into account the fact that the initial
state of the system remains unknown. For instance, the system could
start with an energy density distribution similar to the CGC model,
which has a fairly large eccentricity. Fig.~\ref{fig:dilu} shows that when allowing
the eccentricity to get diluted according to Eq.~(\ref{dilution}),
it takes a time of $\tau\sim 1.5$ fm/c until the eccentricity
has shrunk to that of the Glauber model. This implies that 
even when assuming no particle interactions (no elliptic flow build-up)
for the first stage of the system evolution, one can get
eccentricities which are Glauber-like after waiting for 
a significant fraction of the system life time. Allowing at least
some particle interactions (which is probably more realistic),
one expects some build-up of elliptic flow already in the dilution (or pre-equilibrium)
phase, and therefore dilution (or ``thermalization'') times of $\tau\sim 2$ fm/c seem not 
to be incompatible with the observed final elliptic flow even for non-vanishing
viscosity.

\section{Summary and conclusions}
\label{sec:four}

In this article, we applied conformal relativistic viscous hydrodynamics
to simulate Au+Au collisions at RHIC at energies of $\sqrt{s}=200$ GeV per
nucleon pair. Besides one first-order transport
coefficient (the shear viscosity) in general there are five
second-order transport coefficients in this theory, for which
one would have to supply values. We provided arguments that 
physical observables in the parameter range accessible to hydrodynamics
(low momenta, central to semi-central collisions) do not seem to be 
strongly dependent on specific (reasonable) choices for any of these
second-order coefficients. On the other hand, we do find a pronounced
dependence of the elliptic flow coefficient on the ratio of shear viscosity over
entropy density, which suggests that by combining 
viscous hydrodynamics and experimental data a measurement of the
quark-gluon plasma viscosity may not be futile.
However, we have shown that our ignorance about the precise distribution
of energy density at the earliest stages of a heavy-ion collision
introduces a large systematic uncertainty in the final
elliptic flow of the hydrodynamic model. Adding to this is
the considerable experimental uncertainty pertaining to the removing
of non-flow contributions to the elliptic flow.
For these reasons, we are unable to make precise statements
about the value of the shear viscosity of the quark-gluon plasma
and in particular cannot place a firm lower bound on $\eta/s$.
Indeed, our hydrodynamic models seem to be able to 
consistently describe experimental data for multiplicity,
radial flow and elliptic flow of bulk charged hadrons for a wide
range of viscosity over entropy ratios,
\beq
\frac{\eta}{s}=0.1\pm0.1({\rm theory})\pm0.08({\rm experiment}),
\label{eq:result}
\eeq
where we estimated the systematic uncertainties for both
theory and experiment from the results shown in Fig.~\ref{fig:v2int}. 
We stress that Eq.~(\ref{eq:result}) does not account for
physics not included in our model, such as finite chemical potential,
bulk viscosity, heat flow, hadron cascades, three-dimensional
fluid dynamic effects and possibly many more. Consistent inclusion
of all these may result in changes of the central value and
theory uncertainty in Eq.~(\ref{eq:result}). Nevertheless,
none of the mentioned refinements 
is currently expected to dramatically increase the elliptic
flow coefficient (though some increase may be expected
when e.g. implementing partial chemical equilibrium \cite{Hirano:2005wx}). 
Therefore, we seem to be able to 
exclude viscosities of $\eta/s\gtrsim 0.5$ 
with high confidence, which indicates that the quark-gluon
plasma displays less friction than any other known laboratory 
fluid \cite{Kovtun:2004de,Schafer:2007pr}. 
Other groups have come to similar conclusions 
\cite{Gavin:2006xd,Adare:2006nq,Drescher:2007cd}.

To better quantify the shear viscosity of the quark-gluon plasma
at RHIC calls for more work, both in theory and experiment.
On the theory side, a promising route seems to be the study
of fluctuations and comparing to existing experimental data 
\cite{Adams:2005aw,Gavin:2006xd,Sorensen:2006nw,Alver:2007qw,Vogel:2007yq,
Drescher:2007ax,Trainor:2007ny,Hama:2007dq}. For instance,
it might be interesting to investigate the critical value of $\eta/s$
for the onset of turbulence in heavy-ion collisions and explore possible consequences
of fully developed turbulence \cite{Romatschke:2007eb}.
However, maybe most importantly, 
a more thorough understanding of the earliest stages of a heavy-ion 
collision, in particular thermalization, 
could fix the initial conditions for hydrodynamics and hence
dramatically reduce the theoretical uncertainty in final observables.

Leaving these ideas for future work, we stress that with the advent
of conformal relativistic viscous hydrodynamics at least
the uncertainties of the hydrodynamic evolution itself
now seem to be under control. We hope that this serves
as another step towards a better understanding of the dynamics
of relativistic heavy-ion collisions.
\section*{Note on version 4 of preprint}
An error was found in the freeze out routine used to obtain the original results (see the published erratum [Phys. Rev. C 79, 039903(E) (2009)]).  The error was corrected and the matching to the experimental data was redone.  The only changes are the values of the parameters listed in Table \ref{tab:par}, and slightly revised Figs.~\ref{fig:Pade}, \ref{mult1}, \ref{fig:v2int}, and \ref{fig:tdep}b, which have been updated in this version 4.
\section*{Acknowledgments}
We thank B.~Alver, H.-J.~Drescher, A.~Dumitru, U.~Heinz, 
P.~Huovinen, T.~ Lappi, Y.~Nara, A.~Poskanzer, M.~Rangamani,
K.~Rajagopal,
D.T.~Son, H.~Song, M.~Strickland, D.~Teaney and L.~Yaffe 
for fruitful discussions. 
Furthermore, we would like to thank 
H.-J.~Drescher, A.~Dumitru and Y.~Nara for 
providing their numerical code for the CGC initial conditions;
M.~Laine and Y.~Schr\"oder for providing their tabulated equation of state 
and permission to reproduce their result in Fig.~\ref{figcs2}; 
and A.~Poskanzer for providing tabulated experimental data from the STAR
collaboration. 
The work of ML and PR was supported by the US Department of Energy, grant 
numbers DE-FG02-97ER41014 and DE-FG02-00ER41132, respectively.


\begin{thebibliography}{99}


\bibitem{Adcox:2004mh}
  K.~Adcox {\it et al.}  [PHENIX Collaboration],
  Nucl.\ Phys.\  A {\bf 757} (2005) 184.

\bibitem{Back:2004je}
  B.~B.~Back {\it et al.},
  Nucl.\ Phys.\  A {\bf 757} (2005) 28.

\bibitem{Arsene:2004fa}
  I.~Arsene {\it et al.}  [BRAHMS Collaboration],
  Nucl.\ Phys.\  A {\bf 757} (2005) 1.

\bibitem{Adams:2005dq}
  J.~Adams {\it et al.}  [STAR Collaboration],
  Nucl.\ Phys.\  A {\bf 757} (2005) 102.

\bibitem{Teaney:2000cw}
  D.~Teaney, J.~Lauret and E.~V.~Shuryak,
  Phys.\ Rev.\ Lett.\  {\bf 86} (2001) 4783.

\bibitem{Huovinen:2001cy}
  P.~Huovinen, P.~F.~Kolb, U.~W.~Heinz, P.~V.~Ruuskanen and S.~A.~Voloshin,
  Phys.\ Lett.\  B {\bf 503} (2001) 58.

\bibitem{Kolb:2001qz}
  P.~F.~Kolb, U.~W.~Heinz, P.~Huovinen, K.~J.~Eskola and K.~Tuominen,
  Nucl.\ Phys.\  A {\bf 696} (2001) 197.

\bibitem{Hirano:2002ds}
  T.~Hirano and K.~Tsuda,
  Phys.\ Rev.\  C {\bf 66} (2002) 054905.

\bibitem{Kolb:2002ve}
  P.~F.~Kolb and R.~Rapp,
  Phys.\ Rev.\  C {\bf 67} (2003) 044903.

\bibitem{Policastro:2001yc}
  G.~Policastro, D.~T.~Son and A.~O.~Starinets,
  Phys.\ Rev.\ Lett.\  {\bf 87} (2001) 081601.

\bibitem{Arnold:2003zc}
  P.~Arnold, G.~D.~Moore and L.~G.~Yaffe,
  JHEP {\bf 0305} (2003) 051.

\bibitem{Nakamura:2004sy}
  A.~Nakamura and S.~Sakai,
  Phys.\ Rev.\ Lett.\  {\bf 94} (2005) 072305.

\bibitem{Arnold:2006fz}
  P.~Arnold, C.~Dogan and G.~D.~Moore,
  Phys.\ Rev.\  D {\bf 74} (2006) 085021.

\bibitem{Huot:2006ys}
  S.~C.~Huot, S.~Jeon and G.~D.~Moore,
  Phys.\ Rev.\ Lett.\  {\bf 98} (2007) 172303.

\bibitem{Gagnon:2006hi}
  J.~S.~Gagnon and S.~Jeon,
  Phys.\ Rev.\  D {\bf 75} (2007) 025014
  [Erratum-ibid.\  D {\bf 76} (2007) 089902].

\bibitem{Aarts:2007wj}
  G.~Aarts, C.~Allton, J.~Foley, S.~Hands and S.~Kim,
  Phys.\ Rev.\ Lett.\  {\bf 99} (2007) 022002.

\bibitem{Meyer:2007ic}
  H.~B.~Meyer,
  Phys.\ Rev.\  D {\bf 76} (2007) 101701.


\bibitem{Buchel:2007mf}
  A.~Buchel,
  Phys.\ Lett.\  B {\bf 663} (2008) 286.


\bibitem{CaronHuot:2007gq}
  S.~Caron-Huot and G.~D.~Moore,
  Phys.\ Rev.\ Lett.\  {\bf 100} (2008) 052301.


\bibitem{Teaney:2003kp}
  D.~Teaney,
  Phys.\ Rev.\  C {\bf 68} (2003) 034913.

\bibitem{Romatschke:2007jx}
  P.~Romatschke,
  Eur.\ Phys.\ J.\  C {\bf 52} (2007) 203.

\bibitem{Romatschke:2007mq}
  P.~Romatschke and U.~Romatschke,
  Phys.\ Rev.\ Lett.\  {\bf 99} (2007) 172301.

\bibitem{Chaudhuri:2007qp}
  A.~K.~Chaudhuri,
  arXiv:0708.1252 [nucl-th].

\bibitem{Muronga:2004sf}
  A.~Muronga and D.~H.~Rischke,
  arXiv:nucl-th/0407114.

\bibitem{Chaudhuri:2005ea}
  A.~K.~Chaudhuri and U.~W.~Heinz,
  J.\ Phys.\ Conf.\ Ser.\  {\bf 50} (2006) 251.


\bibitem{Muronga:2005pk}
  A.~Muronga,
  J.\ Phys.\ G {\bf 31} (2005) S1035.


\bibitem{Chaudhuri:2006jd}
  A.~K.~Chaudhuri,
  Phys.\ Rev.\  C {\bf 74} (2006) 044904.


\bibitem{Mota:2007tz}
  Ph.~Mota, G.~S.~Denicol, T.~Koide and T.~Kodama,
  J.\ Phys.\ G {\bf 34} (2007) S1011.

\bibitem{Song:2007fn}
  H.~Song and U.~W.~Heinz,
  Phys.\ Lett.\  B {\bf 658} (2008) 279.




\bibitem{Dusling:2007gi}
  K.~Dusling and D.~Teaney,
  Phys.\ Rev.\  C {\bf 77} (2008) 034905.


\bibitem{Song:2007ux}
  H.~Song and U.~W.~Heinz,
  arXiv:0712.3715 [nucl-th].

\bibitem{IS0a}
W.~Israel, Ann.Phys. {\bf 100} (1976) 310.
\bibitem{IS0b}
W.~Israel and J.M.~Stewart, Phys. Lett. {\bf 58A} (1976)  213.

\bibitem{IS1}
W.~Israel and J.M.~Stewart, Ann.Phys. {\bf 118}, (1979) 341.

\bibitem{Mueller}
I.-Shih Liu, I.~M\"uller and T.~Ruggeri, Ann.\ Phys.\ {\bf 169}
(1986) 191.

\bibitem{Muronga:2003ta}
  A.~Muronga,
  Phys.\ Rev.\  C {\bf 69} (2004) 034903.

\bibitem{Heinz:2005bw}
  U.~W.~Heinz, H.~Song and A.~K.~Chaudhuri,
  Phys.\ Rev.\  C {\bf 73} (2006) 034904.

\bibitem{Baier:2006um}
  R.~Baier, P.~Romatschke and U.~A.~Wiedemann,
  Phys.\ Rev.\  C {\bf 73} (2006) 064903.

\bibitem{Koide:2006ef}
  T.~Koide, G.~S.~Denicol, Ph.~Mota and T.~Kodama,
  Phys.\ Rev.\  C {\bf 75} (2007) 034909.


\bibitem{Baier:2007ix}
  R.~Baier, P.~Romatschke, D.~T.~Son, A.~O.~Starinets and M.~A.~Stephanov,
  JHEP {\bf 0804} (2008) 100.


\bibitem{Maldacena:1997re}
  J.~M.~Maldacena,
  Adv.\ Theor.\ Math.\ Phys.\  {\bf 2} (1998) 231
  [Int.\ J.\ Theor.\ Phys.\  {\bf 38} (1999) 1113].

\bibitem{Bhattacharyya:2008jc}
  S.~Bhattacharyya, V.~E.~Hubeny, S.~Minwalla and M.~Rangamani,
  JHEP {\bf 0802} (2008) 045.

\bibitem{Natsuume:2007ty}
  M.~Natsuume and T.~Okamura,
  Phys.\ Rev.\  D {\bf 77} (2008) 066014.



\bibitem{Kharzeev:2007wb}
  D.~Kharzeev and K.~Tuchin,
  arXiv:0705.4280 [hep-ph].

\bibitem{Sakai:2007cm}
  S.~Sakai and A.~Nakamura,
  PoS {\bf LAT2007} (2007) 221.

\bibitem{Meyer:2007dy}
  H.~B.~Meyer,
  Phys.\ Rev.\ Lett.\  {\bf 100} (2008) 162001.


\bibitem{Karsch:2007jc}
  F.~Karsch, D.~Kharzeev and K.~Tuchin,
  Phys.\ Lett.\  B {\bf 663} (2008) 217.

\bibitem{Muronga:2001zk}
  A.~Muronga,
  Phys.\ Rev.\ Lett.\  {\bf 88} (2002) 062302
  [Erratum-ibid.\  {\bf 89} (2002) 159901].

\bibitem{Baier:2006gy}
  R.~Baier and P.~Romatschke,
  Eur.\ Phys.\ J.\  C {\bf 51} (2007) 677.



\bibitem{Huovinen:2006jp}
  P.~Huovinen and P.~V.~Ruuskanen,
  Ann.\ Rev.\ Nucl.\ Part.\ Sci.\  {\bf 56} (2006) 163.


\bibitem{codedown}
C++ versions of the relativistic viscous hydrodynamic
codes with and without radial symmetry may be
obtained from http://hep.itp.tuwien.ac.at/\verb ~ paulrom/

\bibitem{NR}
Numerical Recipes in C, 2nd edition, Cambridge University Press,
1992.

\bibitem{Dumitru:2007qr}
  A.~Dumitru, E.~Molnar and Y.~Nara,
  Phys.\ Rev.\  C {\bf 76} (2007) 024910.

\bibitem{Aoki:2006we}
  Y.~Aoki, G.~Endrodi, Z.~Fodor, S.~D.~Katz and K.~K.~Szabo,
  Nature {\bf 443} (2006) 675.


\bibitem{Laine:2006cp}
  M.~Laine and Y.~Schroder,
  Phys.\ Rev.\  D {\bf 73} (2006) 085009.

\bibitem{CooperFrye}
F.~Cooper and G.~Frye, Phys.\ Rev.\ D {\bf 10} (1974) 186.

\bibitem{Hung:1997du}
  C.~M.~Hung and E.~V.~Shuryak,
  Phys.\ Rev.\  C {\bf 57} (1998) 1891.

\bibitem{Ruuskanen:1986py}
  P.~V.~Ruuskanen,
  Acta Phys.\ Polon.\  B {\bf 18} (1987) 551.

\bibitem{Rischke:1996em}
  D.~H.~Rischke and M.~Gyulassy,
  Nucl.\ Phys.\  A {\bf 608} (1996) 479.

\bibitem{deGroot}
S.R.de Groot, W.A. van Leeuwen and Ch.G. van Weert, ``Relativistic
Kinetic Theory'', North-Holland Publishing Company (1980).

\bibitem{Kolb:2003dz}
  P.~F.~Kolb and U.~W.~Heinz,
  arXiv:nucl-th/0305084.

\bibitem{Sollfrank:1990qz}
  J.~Sollfrank, P.~Koch and U.~W.~Heinz,
  Phys.\ Lett.\ B {\bf 252} (1990) 256.

\bibitem{Sollfrank:1991xm}
  J.~Sollfrank, P.~Koch and U.~W.~Heinz,
  Z.\ Phys.\ C {\bf 52} (1991) 593.

\bibitem{OSCAR}
AZHYDRO Version 0.2, available from
http://karman.physics.purdue.edu/OSCAR/

\bibitem{Bass:2000ib}
  S.~A.~Bass and A.~Dumitru,
  Phys.\ Rev.\  C {\bf 61} (2000) 064909.

\bibitem{Teaney:2001av}
  D.~Teaney, J.~Lauret and E.~V.~Shuryak,
  arXiv:nucl-th/0110037.

\bibitem{Hirano:2005xf}
  T.~Hirano, U.~W.~Heinz, D.~Kharzeev, R.~Lacey and Y.~Nara,
  Phys.\ Lett.\  B {\bf 636} (2006) 299.

\bibitem{Nonaka:2006yn}
  C.~Nonaka and S.~A.~Bass,
  Phys.\ Rev.\  C {\bf 75} (2007) 014902.


\bibitem{Bhalerao:2007ek}
  R.~S.~Bhalerao and S.~Gupta,
  Phys.\ Rev.\  C {\bf 77} (2008) 014902.


\bibitem{Baym:1984sr}
  G.~Baym, B.~L.~Friman, J.~P.~Blaizot, M.~Soyeur and W.~Czyz,
  Nucl.\ Phys.\  A {\bf 407} (1983) 541.

\bibitem{Chojnacki:2006tv}
  M.~Chojnacki and W.~Florkowski,
  Phys.\ Rev.\  C {\bf 74} (2006) 034905.

\bibitem{Nagy:2007xn}
  M.~I.~Nagy, T.~Csorgo and M.~Csanad,
  Phys.\ Rev.\  C {\bf 77} (2008) 024908.


\bibitem{Kharzeev:2002ei}
  D.~Kharzeev, E.~Levin and M.~Nardi,
  Nucl.\ Phys.\  A {\bf 730} (2004) 448
  [Erratum-ibid.\  A {\bf 743} (2004) 329].

\bibitem{Drescher:2006pi}
  H.~J.~Drescher, A.~Dumitru, A.~Hayashigaki and Y.~Nara,
  Phys.\ Rev.\  C {\bf 74} (2006) 044905.

\bibitem{McLerran:1993ni}
  L.~D.~McLerran and R.~Venugopalan,
  Phys.\ Rev.\  D {\bf 49} (1994) 2233.

\bibitem{McLerran:1993ka}
  L.~D.~McLerran and R.~Venugopalan,
  Phys.\ Rev.\  D {\bf 49} (1994) 3352.

\bibitem{Lappi:2006xc}
  T.~Lappi and R.~Venugopalan,
  Phys.\ Rev.\  C {\bf 74} (2006) 054905.

\bibitem{Kolb:1999it}
  P.~F.~Kolb, J.~Sollfrank and U.~W.~Heinz,
  Phys.\ Lett.\  B {\bf 459} (1999) 667.

\bibitem{Ollitrault:1992bk}
  J.~Y.~Ollitrault,
  Phys.\ Rev.\  D {\bf 46} (1992) 229.

\bibitem{Kovtun:2004de}
  P.~Kovtun, D.~T.~Son and A.~O.~Starinets,
  Phys.\ Rev.\ Lett.\  {\bf 94} (2005) 111601.


\bibitem{Kolb:2000sd}
  P.~F.~Kolb, J.~Sollfrank and U.~W.~Heinz,
  Phys.\ Rev.\  C {\bf 62} (2000) 054909.

\bibitem{Heinz:2004pj}
  U.~W.~Heinz,
  AIP Conf.\ Proc.\  {\bf 739} (2005) 163.


\bibitem{Lublinsky:2007mm}
  M.~Lublinsky and E.~Shuryak,
  Phys.\ Rev.\  C {\bf 76} (2007) 021901.


\bibitem{Adler:2003cb}
  S.~S.~Adler {\it et al.}  [PHENIX Collaboration],
  Phys.\ Rev.\  C {\bf 69} (2004) 034909.

\bibitem{Alver:2007qw}
  B.~Alver {\it et al.}  [PHOBOS Collaboration],
  Int.\ J.\ Mod.\ Phys.\  E {\bf 16} (2008) 3331.

\bibitem{Ollitrault:1995dy}
  J.~Y.~Ollitrault,
  Nucl.\ Phys.\  A {\bf 590} (1995) 561C.

\bibitem{Abelev:2008ed}
  B.I.~Abelev {\it et al.}  [STAR Collaboration],
  arXiv:0801.3466 [nucl-ex].

\bibitem{Poskanzer}
A.~Poskanzer, private communication.

\bibitem{Pasi}
P.~Huovinen, talk given at RHIC Winter Workshop, INT, Seattle, 2002

\bibitem{Broniowski:2008vp}
  W.~Broniowski, M.~Chojnacki, W.~Florkowski and A.~Kisiel,
  arXiv:0801.4361 [nucl-th].

\bibitem{Hirano:2005wx}
  T.~Hirano and M.~Gyulassy,
  Nucl.\ Phys.\  A {\bf 769} (2006) 71
  [arXiv:nucl-th/0506049].

\bibitem{Schafer:2007pr}
  T.~Schafer,
  Phys.\ Rev.\  A {\bf 76} (2007) 063618.


\bibitem{Gavin:2006xd}
  S.~Gavin and M.~Abdel-Aziz,
  Phys.\ Rev.\ Lett.\  {\bf 97} (2006) 162302.

\bibitem{Adare:2006nq}
  A.~Adare {\it et al.}  [PHENIX Collaboration],
  Phys.\ Rev.\ Lett.\  {\bf 98} (2007) 172301.

\bibitem{Drescher:2007cd}
  H.~J.~Drescher, A.~Dumitru, C.~Gombeaud and J.~Y.~Ollitrault,
  Phys.\ Rev.\  C {\bf 76} (2007) 024905.


\bibitem{Adams:2005aw}
  J.~Adams {\it et al.}  [STAR Collaboration],
  J.\ Phys.\ G {\bf 32} (2006) L37.



\bibitem{Sorensen:2006nw}
  P.~Sorensen  [STAR Collaboration],
  J.\ Phys.\ G {\bf 34} (2007) S897.

\bibitem{Vogel:2007yq}
  S.~Vogel, G.~Torrieri and M.~Bleicher,
  arXiv:nucl-th/0703031.

\bibitem{Drescher:2007ax}
  H.~J.~Drescher and Y.~Nara,
  Phys.\ Rev.\  C {\bf 76} (2007) 041903.

\bibitem{Trainor:2007ny}
  T.~A.~Trainor,
  Mod.\ Phys.\ Lett.\  A {\bf 23} (2008) 569.

\bibitem{Hama:2007dq}
  Y.~Hama, R.~Peterson G.Andrade, F.~Grassi, W.~L.~Qian, T.~Osada, C.~E.~Aguiar and T.~Kodama,
  arXiv:0711.4544 [hep-ph].

\bibitem{Romatschke:2007eb}
  P.~Romatschke,
  arXiv:0710.0016 [nucl-th].

\end{thebibliography}
\end{document}